\documentstyle[aps,prb,multicol,epsf,graphicx]{revtex}

\newcommand \bea {\begin{eqnarray}}
\newcommand \eea {\end{eqnarray}}
\newcommand \be {\begin{equation}}
\newcommand \ee {\end{equation}}
\newcommand \eps {\epsilon}
\newcommand \bi {\bibitem}

\begin{document}

\title{The disordered Backgammon model}

\author{L. Leuzzi$^{(1)}$ and F. Ritort$^{(2,3)}$}

\address{(1) Instituut voor Theoretische Fysica\\
        Universiteit van Amsterdam and FOM\\
        Valckenierstraat 65, 1018 XE Amsterdam\\
(2) Departament de F\'{\i}sica Fonamental,
 Facultat de F\'{\i}sica, Universitat de Barcelona\\ Diagonal 647,
 08028 Barcelona,Spain\\
 (3) LPTHE et Universit\'e Pierre et Marie Curie, Paris VI\\
        Universit\'e Denis Diderot, Paris VII\\
        Boite 126, Tour 16, 1$^{\it er}$ \'etage, 4 place Jussieu\\
        F-75252 Paris Cedex 05, France\\
E-Mail:leuzzi@wins.uva.nl,ritort@ffn.ub.es
}


\maketitle
\begin{abstract}
In this paper we consider an exactly solvable model which displays
glassy behavior at zero temperature due to entropic barriers. The new
ingredient of the model is the existence of different energy scales or
modes associated to different relaxational
time-scales. Low-temperature relaxation takes place by partial
equilibration of successive lower energy modes. An adiabatic scaling solution,
 defined in terms of a threshold energy scale $\eps^*$,
is proposed. For such a solution, modes with energy $\eps\gg\eps^*$
are equilibrated at the bath temperature, modes with $\eps\ll\eps^*$
remain out of equilibrium and relaxation occurs in the neighborhood of
the threshold $\eps\sim \eps^*$. The model is presented as a toy
example to investigate conditions related to the existence of an
effective temperature in glassy systems and its possible dependence on
the energy sector probed by the corresponding observable.
\end{abstract}

\vspace{.2cm}
\section{Introduction}
The study of exactly solvable models has been always an active area of
research in the field of statistical physics. They help us to grasp
general principles governing the physical behavior of realistic
systems which, due to the complicated interactions among the different
constituents, cannot be predicted using standard perturbative
techniques.  Glasses in general are systems falling into this
category. The slow relaxation of glasses observed in the laboratory is
a consequence of the simultaneous interplay of  its
constituents that yields a very complex and rich phenomenology.

It is well known that glasses fall out of equilibrium when the
characteristic observation time is larger than their relaxation
time. Because the relaxation time is strongly dependent on 
temperature, it turns out that glasses are immediately out of equilibrium
as soon as the temperature is few degrees below the glass
transition. Well below the glass transition temperature
$T_g$ no time evolution is apparently observed in
the glass and one is tempted to conclude that the glass is in a
stationary state. Nothing more far from the truth. 
Glasses still relax
but slow enough for any change to be observable in laboratory
time-scales. Old experiments on polymers reveal that the slowly relaxing
state corresponds to an aging state \cite{struick}. That is, if the system is perturbed
while being in its aging state, then the characteristic time associated
to the response of the system scales with the age of the system
(i.e. the time elapsed since it was quenched). Another way to look at
this aging phenomena is to evaluate the time autocorrelation
function. It is observed that the typical decorrelation time scales
with the age of the system \cite{AGING}. 

\begin{figure}[hbp!]
\begin{center}
\rotatebox{270}
{\includegraphics*[width=6cm,height=10cm]{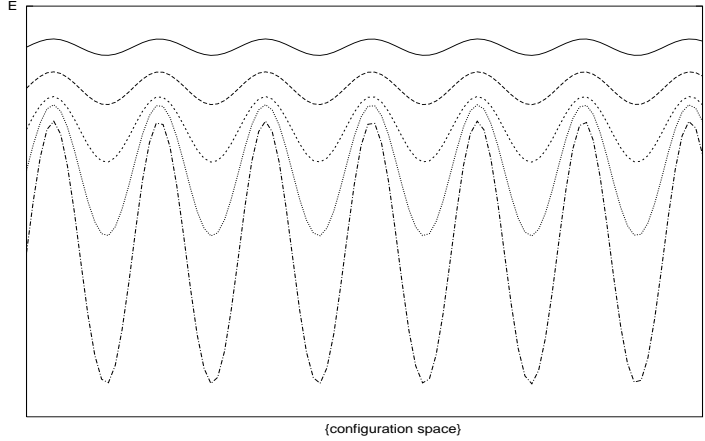}}
\vskip 0.05in
\caption{A one dimensional  example of an energy landscape 
of collective modes: high energy collective modes are separated by
low barriers while low energy collective modes are separated by high
barriers.\label{figintro}}
\end{center}
\end{figure}

A simple scenario to explain these results is the following. Consider
a liquid well above $T_g$ where correlations decay exponentially with
 time. One may consider the resultant behavior of the liquid as the
superposition of different and independent harmonic modes.  Each of
this energy modes corresponds to a normal mode of a system and
describes a collective oscillation of $N$ atoms around their local
minimum.  This is the harmonic approximation which is known to work
quite well in liquids. Nevertheless, already  as the temperature goes below a
{\em critical} temperature $T_c$ (the transition temperature predicted
by the mode-coupling theory\cite{goetze}) other collective modes
different from the standard vibrational ones become important.  The
nature of these modes is quite different from the usual harmonic
normal modes because they do not represent oscillations around a given
configuration within a metastable well but transitions among different
wells.  
These modes are reminiscent of some type of instanton
solutions recently computed in the framework of some spin glass models\cite{inst}.  
Below $T_c$ relaxational dynamics proceeds by
activation over the barriers characterizing these collective
modes. Now, the main difference between these collective modes and the
usual harmonic modes relies on how they relax to equilibrium when put
in contact with a thermal bath at temperature $T$. Relaxation to
equilibrium is determined by the height of the energy barriers
separating different modes. Suppose a given normal mode has frequency
$\omega_k$ and energy $E_k\propto \omega_k^2$. The relaxation time for
each of these modes is typically of order $\tau_k\sim \exp(E_k/K_BT)$.
Therefore, as the energy $E_k$ is lower than the thermal bath
temperature this mode rapidly equilibrates.  On the contrary, if
$E_k~\gg~K_B T$, this mode remains frozen. Collective modes are
different.  
As the reference energy of the collective modes deplete,
the typical barrier separating these modes increases leading to the
contrary behaviour and to super-activation effects. 
While high energy
collective modes are separated by low barriers, 
 low energy
collective modes are separated by high barriers. A simple schematic
representation of this scenario in a one dimensional configurational
space is shown in figure~\ref{figintro}. This behavior is common to
the majority of exactly solvable glassy models \cite{pspin,RPRL95,BPR} ,
phenomenological trap models \cite{trap} and kinetically
constrained models \cite{FacDyn} to cite a few.

In what follows we will use the generic word {\em mode} to refer to this
kind of collective excitation.  Let us label the modes with the integer
variable $r$ and let us denote their energy by $\eps_r$.  Let us suppose
that the energy levels are ordered from lower to higher energies
according to the label $r$.  It is natural to assume that there
is a characteristic mode $r^*$ with associated energy $\eps^*$ such
that, all modes with $\eps_r\gg \eps^*$ have already relaxed, while in
the other limit, $\eps_r\ll \eps^*$, all modes are frozen.  If the
system is quenched well below $T_c$ then equilibration cannot be
achieved in laboratory time scales, this means that all modes below
$\eps^*$ remain frozen while modes above $\eps^*$ remain equilibrated at
the bath temperature.  The energy threshold $\eps^*$ decays with time
because, as time goes by, higher barriers are accessible to the
system. The resulting scenario is that of a liquid where collective
modes above $\eps^*$ are in some sort of {\em local equilibrium} at the
temperature of the bath $T$ while modes below $\eps^*$ are frozen.  This
scenario, as it stands, is too naive because it is based on the
assumption that there are no dynamical correlations between the
different modes, i.e. high energy modes do not influence low energy
modes. While this is true in the equilibrium state it may not be valid
(an indeed it is not) when any type of local dynamics induces
correlations between different modes.  Still it is interesting to
investigate in which conditions such a description turns out to be
correct.  Examples where these type of description holds are mean-field
models\cite{pspin,KirkThiru,DRPA92,KGRCMPA94,CKPRL93}.  By mean-field we
mean those models where there is no spatial dimensionality associated
with the set of interactions. In these cases, different modes are not
related to different length scales. Therefore local dynamical rules do
not necessarily induce correlations between the different modes.
Generally speaking, the identification of this energy scale remains an
open problem for which we do not have yet a complete understanding.

This energy scale $\eps^*$ is related to what has received the name of
effective or fictive temperature $T_{\rm eff}$ in the most recent
literature about glassy dynamics
\cite{TeffCKP,FRJPA97,THEO1,THEO2,FVJPA00,luck}.  $T_{\rm eff}$ is an
effective time-dependent parameter describing equilibrium fluctuations
for those thermalized modes with $\eps\gg\eps^*=k_BT_{\rm eff}$.  While
a thermometer coupled to the fast modes is expected to measure the bath
temperature, a thermometer coupled to the slower ones should yield the
effective temperature. Hence, the effective temperature corresponds to
an energy threshold which separates collective modes which are frozen
from those which have relaxed to the bath temperature.  Relaxation, then,
takes place at energy scales in the neighborhood of the threshold value
$\eps^*$. For real systems this energy threshold is related to the
typical volume of the drop which is able to release strain energy during
its relaxation to equilibrium. A scenario on this type of physical
mechanism for glassy relaxation has been recently introduced in
\cite{CR1} where this threshold energy $\eps^*$ has been related to the
size of the cooperative region as an explanation of the {\em fragility}
and super-activation anomalies in real glasses.

The existence on an effective temperature is tightly related to the
validity of some approximations used in the context of slowly relaxing
systems such as the adiabatic approximation. Within the adiabatic
approximation one obtains a Markovian description for the
dynamics. Again, this Markovian description encodes within a single
parameter (the effective temperature) all the complicated previous past
of the system. The adiabatic approximation has been shown to give the
correct asymptotic dynamical behavior in some simple models of glasses
such as the Backgammon model \cite{FREPL95,BGM} 
while for the majority of the other most
famous models in the literature (for instance the p-spin model\cite{pspin}) 
the implementation of such approximation
is generally not yet known.  Quite probably a Markovian description for glassy
 dynamics 
is unrealistic and the original idea of experimentalists \cite{TOOL}
to encode the dynamical behavior into effective parameters such as the
effective pressure or the effective (else said fictive) 
temperature could be only labels
without a deep physical meaning.

These ideas have been contrasted in several exactly solvable models. All
these models have the advantage of being mean-field, hence dynamical
equations can be closed in one or another way. Still, the mean-field
character of these models does not generally allow to investigate in a
simple way the relaxational properties of the different modes of the
spectrum.  In general, the dynamics of all these models has been studied
by addressing the closure of the dynamics associated to global
quantities (such as the energy or correlation-response functions) which
do not discern the contribution to the global relaxation of the
different energy modes.  Examples of these models are the $p$-spin model
\cite{pspin,KirkThiru,CKPRL93,THEO1}, the Backgammon
\cite{luck,FREPL95,BGM,FRJSP96} the harmonic oscillator and other
spherical spin models \cite{BPR,THEO1,THEO2}.  The question of the
existence of an effective temperature is also tightly related to the
particular way in which the fluctuation-dissipation theorem (FDT) is
violated \cite{CKPRL93,Somp,BB}.  The question whether there exists a
single effective temperature describing the violation of FDT is still
controversial (see, for instance, \cite{SF}). We believe that this and
other related questions can be better addressed, even at the level of
mean-field models, by analyzing the contribution of all different modes.
The analysis of the harmonic vibrational modes in glasses is already an
interesting step in this direction \cite{CPG} although here we have in
mind other type of collective modes.

The purpose of this paper is the study of a new model where the
dynamical relaxation of the different energy modes can be made
explicitly clear. This model is what we refer to as the disordered
Backgammon model (DBG model) and consists in a generalization of the
Backgammon model to allow different energies for different boxes. Again,
just like its predecessor, the slow relaxation of this model is due to
entropic barriers.  For the DBG model we show the existence of an energy
threshold $\eps^*(t)$ which separates equilibrated form non-equilibrated
modes. The DBG model (with its associated threshold $\eps^*(t)$)
provides a microscopic realization that is reminiscent of some
phenomenological models proposed in the past such as the trap model in a
tree considered by Bouchaud and Dean \cite{BD}. The advantage in the DBG
model is that now one can exhaustively investigate the distinct
relaxation of each of the different energy modes verifying whether the
scenario of the effective temperature presented before holds. This will
help to better understand the meaning of the effective temperature and
its relation with the violation fluctuation-dissipation theorem.

In section II we introduce the disordered Backgammon (DBG) model. Its
thermodynamic properties are reported in section III, where the special
case of very low temperatures is explicitly worked out.  In section IV
we present the dynamical equations whose solutions are found within a
new type of adiabatic approximation in section V.  There the features of
such an approximation are carefully analyzed.  In section VI numerical
results are presented for two specific models belonging to the family of
DBG models .  Finally, in section VII, a novel method is introduced in
order to estimate the threshold energy scale $\eps^*$ directly from the
dynamics.  In section VIII we present our conclusions. Some technical
issues are presented in four different appendices.

\section{The disordered Backgammon  model}

\subsection{Definition of the model}

Let us take $N$ particles which can occupy $N$ boxes, each one
labeled by an index $r$ which runs from $1$ to $N$. Suppose now that
all particles are distributed among the boxes. A given box
$r$ contributes to the Hamiltonian 
with an energy $-\eps_r$ only when it is empty. In this
case the total Hamiltonian of the system reads

\be
{\cal H}=-\sum_{r=1}^N\eps_r\delta_{n_r,0} \ ,
\label{eq0}
\ee

\noindent
where $\delta$ is the Kronecker delta and $n_r$ denotes the occupancy
or number of particles in box $r$. The $\eps_r$ are quenched random
variables extracted from a distribution $g(\eps)$ which we assume to
be defined only for $\eps\ge 0$. The interest of this definition will
be discussed below.  Like in the original Backgammon (BG) model
\cite{RPRL95} we consider Monte Carlo mean-field dynamics where a
particle is randomly chosen in a departure box and a move to an
arrival box $a$ is proposed. If $d$ denotes the departure box, the
proposed change is accepted according to the Metropolis rule with
probability $W(\Delta E)={\rm Min}(1,\exp(-\beta \Delta E))$ where
$\Delta E=\eps_a\delta_{n_a,0}-\eps_d\delta_{n_d,1}$.  Note that the
departure box satisfies $n_d\ge 1$ and departure boxes are chosen with
probability $n_d/N$. This dynamics corresponds to Maxwell statistics
where particle are distinguishable and differs from the one
corresponding to Bose statistics \cite{COM,BGM,PRADOS,luck} where particles are
indistinguishable and arrival boxes are chosen with uniform
probability $1/N$.

In the dynamics the total number of particles is conserved 
so that  the occupancies satisfy the closure condition

\be
\sum_{r=1}^N n_r=N~~~~~.
\label{eq1}
\ee

Now we want to show that the interesting case corresponds to the
situation where $g(\eps)$ is only defined for $\eps\ge 0$. 
In this
case, the dynamics turns out to be extremely slow at low temperatures,
similarly to what happens for the original BG model. The difference
lies in the type of ground state. The ground state of (\ref{eq0})
corresponds to the case where all particles occupy a single box, the
one with the smallest value of $\eps$. Let us denote by $\eps_0$ this
smallest value. Then the ground state energy is given by the relation
$E_{GS}=-\sum_{r=1}^N\eps_r+\eps_0$. Since all the $\eps$ are
positive no other configuration can have a lower energy. If 
$g(\eps)$ is a continuous distribution the ground state is also
unique. Now it is easy to understand that, during the dynamical
evolution at zero temperature, all boxes with high values of $\eps$
become empty quite soon and the dynamics involves boxes with
progressively lower values of $\eps$. The asymptotic dynamics is then
determined by the behavior of the distribution $g(\eps)$ in the limit
$\eps\to 0$.  If $g(\eps)~\sim~\eps^{\alpha}$, for $\eps\to 0$, we
will show that the asymptotic long-time properties only depend on
$\alpha$. Note that the normalization of the $g(\eps)$ imposes $\alpha>-1$. This
classification includes also the original BG model where there is no
disorder at all. In that case $g(\eps)=\delta(\eps-1)$ so the
distribution has a finite gap at $\eps=0$. The behavior corresponding
 to this singular energy distribution
can be obtained from the previous one in the limiting case $\alpha\to
\infty$.

One important aspect of the model (\ref{eq0}) is that, in the presence
of disorder, it is not invariant under an arbitrary constant shift of
the energy levels.  Actually, by changing $\eps_r\to\eps'_r=\eps_r+c$
with $c\ge0$, the model turns out to be a combination of the original
model plus the classical BG model.
After shifting, the new distribution
$g(\eps'-c)$ has a finite gap (equal to $c$ plus the gap of the original
distribution). The new model corresponds again to the $\alpha\to\infty$
case and the asymptotic dynamical behavior coincides with that of the
standard BG model. As we will see later the present model is
characterized by an energy threshold $\eps^*$ which drives relaxation to
the stationary state. Only when the energy threshold can go to zero we
have a different asymptotic behavior. For all models with a finite gap,
$\eps^*$ cannot be smaller than the gap, hence asymptotically sticks to
the gap and the relaxational behavior of the DBG model with a finite gap
corresponds to
that of the standard BG model.

One of the outstanding features of this model is that a description
of the dynamics in the framework of an adiabatic approximation turns out
to be totally independent from the type of distribution $g(\eps)$ (and hence on
$\alpha$) despite of the fact that the asymptotic long-time behavior of
the effective temperature and of the internal 
energy depend on the value of $\alpha$.

\subsection{Observables}

Like in the original BG model we define the occupation probabilities,
$P_k$, that a box contains $k$ particle,

\be
P_k=\frac{1}{N}\sum_{r=1}^N\delta_{n_r,k} \ ,
\label{defP}
\ee

\noindent
and the corresponding densities that a box of energy $\eps$ contains $k$ 
particles

\bea
g_k(\eps)&=&\frac{1}{N}\sum_{r=1}^N\delta(\eps_r-\eps)\delta_{n_r,k}~~~~k\ge0 \ ;
\label{defgk}
\\
g(\eps)&=&\frac{1}{N}\sum_{r=1}^N\delta(\eps_r-\eps) \ .
\label{defg}
\eea

The $P_k$ and the $g_k$ are related by

\be
P_k=\int_0^{\infty}g_k(\eps)d\eps~~~~~k\ge 0 \ ,
\label{eqPg}
\ee
\noindent and the conservation of particles reads

\be
\sum_{k=0}^{\infty}P_k=1~~~~;~~~~
\sum_{k=0}^{\infty}g_k(\eps)=\frac{g(\eps)}{\int_0^{\infty}g(\eps)d\eps}=g(\eps) \ .
\label{eq5}
\ee

The energy can be expressed in terms of the density $g_0(\eps)$ as

\be
E=-\int_0^{\infty}d\eps~\eps~ g_0(\eps)~~~.
\label{eq6}\ee

This set of observables depend on time through the time evolution of the
occupancies $n_r$ of all boxes. In the next section we analyze the main
equilibrium properties of the model.

\section{Equilibrium behavior}

The solution of the thermodynamics proceeds similarly as
for the case of the original BG model. The partition function can be
computed in the grand partition ensemble. It reads

\be
{\cal Z}_{G C}=\sum_{N=0}^{\infty}{\cal Z}_C(N)z^N \ ,
\label{eqb1}
\ee
\noindent 
where $z=\exp(\beta \mu)$ is the fugacity, $\mu$ is the chemical potential
 and ${\cal Z}_C(N)$ stands for
the canonical partition function of a system with $N$ particles. The
canonical partition function can be written as

\be
{\cal Z}_{C}=\sum_{n_r=0}^{N}\frac{N!}{\prod_{r=1}^N n_r!}
\exp(\beta\sum_{r=1}^N\eps_r\delta_{n_r,0})\delta_{N,\sum_{r=1}^N n_r}~~~,
\label{eqb2}
\ee

\noindent
where $\delta_{i,j}$ is the Kronecker delta. Introducing this
expression in (\ref{eqb1}) we can write down ${\cal Z}_{G C}$ as an
unrestricted sum for all the occupancies $n_r$

\be
{\cal Z}_{G C}=N!\sum_{n_r=0}^{\infty}
\prod_{r=1}^N\frac{z^{n_r}}{n_r!}\exp(\beta\eps_r\delta_{n_r,0}) \ .
\label{eqb3}
\ee

The factor $N!/\prod_r n_r!$ in the partition function ${\cal Z}_{C}$ is
introduced to account for the distinguishability of particles. This
factor leads to an over-extensive entropy (i.e., the  Gibbs paradox)
which can be cured eliminating from ${\cal Z}_{G C}$ the over-counting
term $N!$ in the numerator. The final result is

\be
{\cal Z}_{G C}=\exp\left\{\sum_{r=1}^N\log\left[\sum_{n=0}^{\infty}
\frac{z^{n}}{n!}\exp\left(\beta\eps_r\delta_{n,0}\right)\right]\right\} \ ,
\label{eqb4}
\ee
yielding the grand-canonical potential energy per box

\be
G=F-\mu=-T\frac{\log({\cal Z}_{G C})}{N}=-T\sum_{r=1}^N
\log\left[\exp(\beta\eps_r)+\exp(z)-1\right]=-T\int_0^{\infty}
g(\eps)\log\left[\exp(\beta\eps)+\exp(z)-1\right]~d\eps~~,
\label{eqb5}
\ee

\noindent
where $F$ is the Helmholtz free-energy per box. The fugacity $z$ is
determined by the conservation condition (\ref{eq1}) which reads 

\be \frac{\partial G}{\partial \mu}=-1~~~{\rm
or,~~~~equivalently, }~~~~\frac{\partial F}{\partial \mu}=0 \ ,
\label{eqb6}
\ee

\noindent
yielding the closure condition

\be 
\int_0^{\infty}
\frac{g(\eps)}{\exp(\beta\eps)+\exp(z)-1}d\eps=\frac{1}{z\exp(z)}~~~~.
\label{eqb7}
\ee

This equation gives the fugacity $z$ as function of $\beta$ and, from
(\ref{eqb5}) and its derivatives, the whole thermodynamics. In
particular, the equilibrium expressions for $g_k(\eps)$ are

\be
g_k^{\rm eq}(\eps)=\frac{z^k g(\eps)\exp(\beta\eps\delta_{k,0})}
{k!(\exp(\beta\eps)+\exp(z)-1)} \ ,
\label{eq7g}
\ee

\noindent
the corresponding $P_k$ being given by eq.(\ref{eqPg}) which, together
with the closure relation (\ref{eqb7}), leads to the expression

\be
P_k^{\rm eq}=\delta_{k,0}(1-\frac{\exp(z)-1}{z\exp(z)})+
(1-\delta_{k,0})\frac{z^{k-1}}{k!\exp(z)} ~~~~.
\label{eq7h}
\ee
Starting from eq.(\ref{eq6}) the equilibrium energy density is obtained as

\be
E^{\rm eq}=-\int_0^{\infty}d\eps\frac{\eps~ g(\eps)\exp(\beta\eps)}
{\exp(\beta\eps)+\exp(z)-1} \ .
\label{eq7e}
\ee

All these
expressions can be evaluated at finite temperature. Note that, although
the values of $P_k^{\rm eq}$ in (\ref{eq7h}) are independent of the disorder
distribution $g(\eps)$, they directly depend on that distribution through
the equilibrium value of $z$ (which obviously depends on the $g(\eps)$).
Of particular interest for the dynamical behavior of the model are the
low-temperature properties that we analyze below.

\subsection{Thermodynamics at low temperatures}

A perturbative expansion can be carried out close to $T\to0$ to find 
the leading behavior of different thermodynamic quantities.
Let us start analyzing the closure condition (\ref{eqb7}). 
Doing the transformation $\eps'=\beta\eps$, eq.(\ref{eqb7}) can be 
rewritten as

\be 
T\int_0^{\infty}
\frac{g(T\eps')}{\exp(\eps')+\exp(z)-1}d\eps'=\frac{1}{z\exp(z)}~~~~.
\label{eqb8}
\ee

In the limit $T\to 0$ the fugacity $z$ depends on the behavior of 
$g(T\eps')$ in the limit $T=0$, i.e. on the behavior of $g(\eps)$ for
$\eps\to 0$. 
Assuming $g(\eps)\sim \eps^{\alpha}$ for $\eps\ll 1$ 
we define the function $h(\eps)$ through the relation 
$g(\eps)=\eps^{\alpha}h(\eps)$, where  $h(\eps)$ is a smooth
function of $\eps$ with $h(0)$ finite. The integral can 
be expanded around $T=0$ by taking successive derivatives of the
function $h$:

\be 
z\exp(z)T^{\alpha+1}\int_0^{\infty}\frac{\eps'^{\alpha}}
{\exp{\eps'}+\exp(z)-1}\sum_{k=0}^{\infty}\frac{h^{(k)}(0)(T\eps')^k}{k!}d\eps'=1 \ .
\label{eqb9}
\ee
Using the asymptotic result $z\to\infty$, when  $T\to 0$, everything reduces
to estimate the following integral in the large $z$ limit:

\be
\int_0^{\infty}dx\frac{x^{\alpha+k}}{\exp(x)+\exp(z)-1}\sim z^{\alpha+k+1}\exp(-z) \ .
\label{eqb10}
\ee

\begin{figure}[tbp]
\begin{center}
{\includegraphics*[width=8cm,height=6cm]{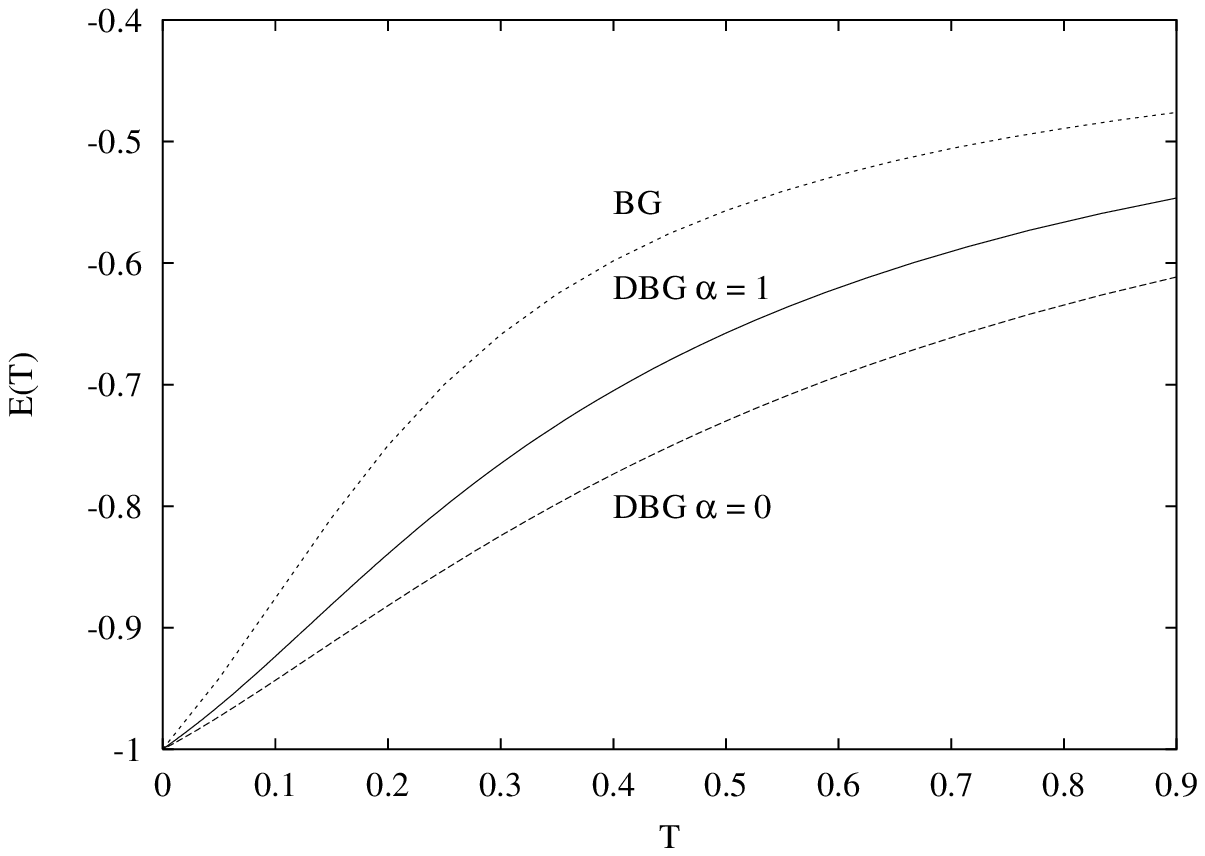}
\includegraphics*[width=8cm,height=6cm]{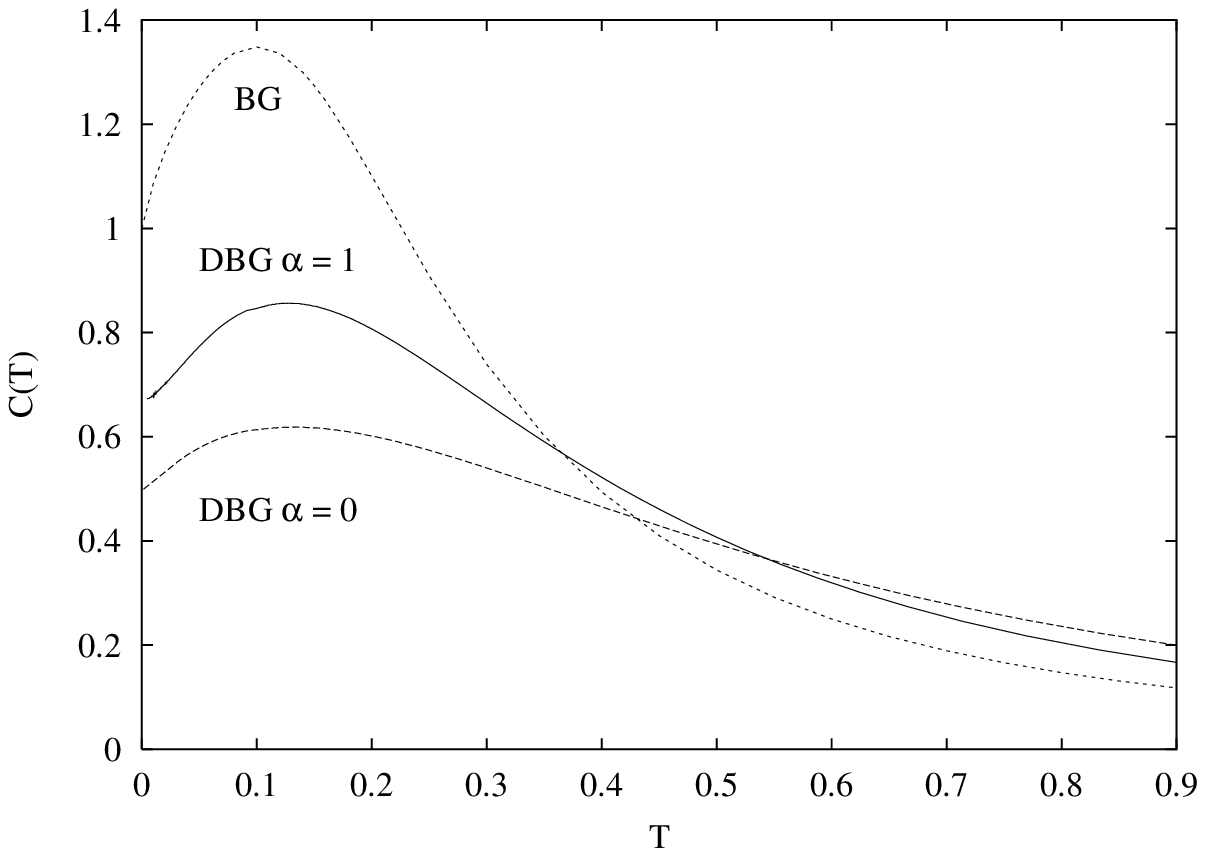}}
\vskip 0.05in
\caption{Left: energy dependence on temperature at equilibrium for 
two DBG models, with $\alpha=1$ and $\alpha=0$. 
Also the energy for the standard BG model ($\alpha\to\infty$) is plotted. 
Right: specific heat for the same models. For all cases (disordered or not) 
it turns out to be  finite at $T=0$.\label{fig0}}
\end{center}
\end{figure}

The term $k=0$ in the series yields the leading behavior for $z$, which
turns out to be

\be
z\sim \beta^{\frac{\alpha+1}{\alpha+2}}
\label{eqbz}
\ee
In a similar way the energy can be computed to leading order in $T$,
\be
E=E_{G S}+a T+{\cal O}(T^2) \ ,
\label{eqb11}
\ee
\noindent giving a finite specific heat at low temperatures.  In section
VI we will show explicitly such a behaviour for two specific DBG models,
one with $\alpha=1$ (defined in eq. (\ref{eqf2}), sec. VI) and the other
with $\alpha=0$ (eq. (\ref{eqf3}), sec. VI).  Solving eq.  (\ref{eqb7})
numerically for $z(T)$ in each specific model and inserting $z(T)$ in
the expression (\ref{eq7e}) for the equilibrium energy density, we get
the energy dependence on the temperature.  In all cases, at equilibrium,
the energy is linear for low $T$, as predicted in eq. (\ref{eqb11}). It
yields, therefore, a finite specific heat as corresponds to a classical
model with Maxwell-Boltzmann statistics (see fig. (\ref{fig0})).

\section{Dynamical equations}

Here we consider the dynamical equations for the occupation
probabilities $P_k$ and their associated densities $g_k(\eps)$.  The
dynamical equations in this model are derived in a similar way as for
the standard BG model. The main difference is that in the DBG the equations for
the occupancies probabilities $P_k$ do not generate a closed hierarchy
of equations. Only for $T=0$ such  a closed  hierarchy is
obtained. As we will see later, this has important consequences when we
become interested in the zero-temperature relaxation.

A hierarchy of equations can only be obtained at the level of the occupation
probability densities $g_k(\eps)$. A detailed derivation of these equations is
reported in the Appendix A. Here we show the final result,

\bea
\frac{\partial g_0(\eps)}{\partial t}&=&
g_1(\epsilon)\left[1+
\int_\epsilon^{\infty}d\epsilon' g_0(\epsilon')\left(
 e^{-\beta(\epsilon'-\epsilon)}-1\right)
\right]
\label{eqg0} \\
\nonumber
&-&g_{0}(\epsilon) 
\left[e^{-\beta\epsilon}+P_1\left(1-e^{-\beta\epsilon}\right)
+\int_0^\epsilon d\epsilon' g_1(\epsilon')\left(
 e^{-\beta(\epsilon-\epsilon')}-1\right)\right]
\hspace*{2 cm} k=0
\ ,
\\
\frac{\partial g_1(\eps)}{\partial t}&=&2g_{2}(\epsilon)\left(
1+\int_0^{\infty}d\epsilon g_0(\epsilon) e^{-\beta\epsilon}-P_0\right)
-g_1(\epsilon)\left[2+
\int_\epsilon^{\infty}d\epsilon' g_0(\epsilon')\left(
 e^{-\beta(\epsilon'-\epsilon)}-1\right)
\right]
\label{eqg1}
\\
\nonumber
&+&g_{0}(\epsilon) 
\left[e^{-\beta\epsilon}+P_1\left(1-e^{-\beta\epsilon}\right)
+\int_0^\epsilon d\epsilon' g_1(\epsilon')\left(
 e^{-\beta(\epsilon-\epsilon')}-1\right)\right]
\hspace*{2cm} k=1 \ ,
\\\frac{\partial g_k(\eps)}{\partial t}&=&(k+1)g_{k+1}(\epsilon)\left(
1+\int_0^{\infty}d\epsilon g_0(\epsilon) e^{-\beta\epsilon}-P_0\right)
\label{eqgk}
\\
\nonumber
&-&g_k(\epsilon)\left[1+k+k\left(
\int_0^{\infty}d\epsilon g_0(\epsilon) e^{-\beta\epsilon}-P_0\right)
\right]+g_{k-1}(\epsilon) \hspace*{3 cm} k>1 \ .
\eea
The equations for the $P_k$ are directly obtained  by integrating the 
$g_k(\eps)$ according to (\ref{eqPg}). The result is

\be
\frac{d P_k(t)}{d t}=
(k+1)\left(P_{k+1}(t)-P_{k}(t)\right)+P_{k-1}
+\left(\int_0^{\infty}d\epsilon g_0(\epsilon) e^{-\beta\epsilon}-P_0\right)
\left[\delta_{k,1}-\delta_{k,0}-k P_k(t)+(k+1)P_{k+1}(t)\right]
\label{eqc2}
\ee
\noindent
with $P_{-1}=0$.
It is easy to check that the equilibrium solutions (\ref{eq7g}) are
indeed stationary solutions.  As previously said, for general $\beta$,
the equations for the $P_k$ do not generate a hierarchy by themselves
but depend on the $g_k(\eps)$ through the distribution $g_0(\eps)$ in
(\ref{eqc2}).  Nevertheless, a remarkable aspect is that they generate a
well defined hierarchy at $T=0$ which coincides with the equations of
the original BG model\cite{FRJSP96}. These are

\be
\frac{d P_k(t)}{d t}
=(k+1)\left(P_{k+1}(t)-P_{k}(t)\right)+P_{k-1}
-P_0
\left[\delta_{k,1}-\delta_{k,0}-k P_k(t)+(k+1)P_{k+1}(t)\right] \ .
\label{eqc3}
\ee

It is easy to understand 
why at $T=0$ the dynamical equations are independent on the
density of states $g(\eps)$. The reasoning is as follows.
For $T=0$, all moves of particles between
 departure and arrival boxes with different energies $\eps_d$ and
$\eps_a$ depend on the precise values of these energies only when the
departure box contains a single 
particle and the arrival box is empty. 
But such a
move does not lead to any change in any of the $P_k$, hence dynamical
equations for the $P_k$ remain independent of $g(\eps)$. Obviously this
does not hold for other observables such as the energy $E$ (i.e. the
mean value of $\eps$ over the density $g_0(\eps)$) 
and  higher moments of
 $g_k(\eps)$. 

This observation is crucial, since we look at the glassy behaviour at $T=0$.
 It turns out that the analysis of the dynamical
equations for the DBG at $T=0$ decomposes into two parts. On the one
hand, the equations for the $P_k$ coincide with those of the original
non-disordered BG model. On the other hand, in order 
 to analyze the behavior of
the energy one must analyze the behavior of the hierarchy of equations
for the $g_k(\eps)$ which is quite complicated. In the next section we
will see how the analysis of these equations can be done within the
framework of a generalized adiabatic approximation.


\section{The adiabatic approximation}

\subsection{Standard adiabatic solution for the $P_k$}

In this section we are interested in the solution of the dynamical
equations at $T=0$. We already saw that the equations (\ref{eqc3}) for the 
$P_k$
coincide with those of the original non-disordered  BG model.
Consequently, the same adiabatic approximation used for the $P_k$ in the
original BG model is still valid for the DBG. Let us remind the main
results \cite{FREPL95,BGM}. 
The key idea behind that approximation is that, while $P_0$
constitutes a slow mode, the other $P_k$ with $k>0$ are fast modes. Hence
they can be considered as if they were in equilibrium at the hyper-surface
in phase space $P_0={\rm constant}$, this constant being given by the
actual value of $P_0$ at time $t$. 
In the original BG model  $P_0=-E$, hence thermalization of the fast
modes $P_k$ ($k>0$) occurs on the hyper-surface of constant energy. For
the DBG model this is not true, the hyper-surface where equilibration of
fast modes occurs does not coincide with the constant energy hyper-surface
simply because the energy and $P_0$ are different quantities. 
Indeed,
we will see later that their leading time behavior is different.

At $T=0$ the
equation for $P_0$ (eq. (\ref{eqc3}) for $k=0$) reads

\be
\frac{\partial P_0}{\partial t}=P_1(1-P_0) \ .
\label{eqd2}
\ee

If local equilibrium is reached on the hyper-surface of constant $P_0$
we can relate $P_1$ to $P_0$ using eqs.(\ref{eq7h}). The simplest way
of dealing with (\ref{eqd2}) is to relate both $P_1$ and $P_0$ to the
time-dependent fugacity $z^*$ writing down a dynamical equation for $z^*$.
Using eq.(\ref{eq7h}) this yields

\be
\frac{\partial z^*}{\partial t}=\frac{z^*(\exp(z^*)-1)}{\exp(z^*)(\exp(z^*)-z^*-1)} \ .
\label{eqd3}
\ee

In the large time limit $z^*$ diverges and the leading asymptotic behavior
is given by $z^*\simeq\log(t)+\log(\log(t))$. The occupation probabilities
are, then, given by the relations (\ref{eq7h}) replacing $z$ by the
time-dependent fugacity $z^*$, as long as  $z^*\gg 1$,

\be
P_k\simeq\delta_{k,0}\left(1-\frac{1}{z^*}\right)+
(1-\delta_{k,0})\frac{(z^*)^{k-1}}{k!\exp(z^*)} ~~~~.
\label{eqd3b}
\ee

Hence, according to (\ref{eqbz}), the inverse
effective temperature is given by the relation

\be
\beta_{\rm eff}\sim (z^*)^{\frac{\alpha+2}{\alpha+1}}\sim (\log(t))^{\frac{\alpha+2}{\alpha+1}}
\label{eqd4}
\ee
\noindent 
and the effective temperature depends on the properties of the disorder
distribution $g(\eps)$ in the limit $\eps\to 0$ through the value of the exponent
$\alpha$. Clearly, when the density of levels decreases as we approach 
$\eps=0$, the relaxation turns out to be slower; the limiting
case being the original BG model for which $\alpha\to\infty$ and
$\beta_{\rm eff}\sim \log(t)$. In the other limit $\alpha\to -1$, when
disorder becomes unnormalized, the inverse 
effective temperature diverges very fast.
Already from (\ref{eqb11}) one can
anticipate that the same asymptotic behavior holds for the energy (see eq. 
(\ref{eqe5}), hence
$\alpha$ interpolates between fast relaxation ($\alpha=-1$) and
very slow relaxation ($\alpha=\infty$). 
A relaxation slower 
than logarithmic is
not possible in the present model.

\subsection{Generalized adiabatic solution for the $g_k(\eps)$}

The equations for the $g_k(\eps)$ at $T=0$ are

\bea
\label{eqd1}
\frac{\partial g_0(\eps)}{\partial t}&=&
g_1(\epsilon)\left[1-
\int_\epsilon^{\infty}d\epsilon' g_0(\epsilon')
\right]
-g_{0}(\epsilon) 
\int_\epsilon^\infty d\epsilon' g_1(\epsilon')
\hspace*{ 3.5 cm} k=0 \ ,
\\
\frac{\partial g_1(\eps)}{\partial t}&=&2g_{2}(\epsilon)\left(
1-P_0\right)
-g_1(\epsilon)\left[2-
\int_\epsilon^{\infty}d\epsilon' g_0(\epsilon')
\right]+g_{0}(\epsilon) 
\int_\epsilon^\infty d\epsilon' g_1(\epsilon')
\hspace*{ 1 cm}k=1 \ ,
\label{eqd1b}
\\
\label{eqd1c}
\frac{\partial g_k(\eps)}{\partial t}&=&(k+1)g_{k+1}(\epsilon)\left(
1-P_0\right)
-g_k(\epsilon)\left[1+k\left(1-P_0\right)
\right]+g_{k-1}(\epsilon) \hspace*{ 2.5 cm} k>1 \ .
\eea

To solve the dynamical equations for the $g_k(\eps)$ in the adiabatic
approximation we note that, contrarily to the global quantities $P_k$,
they cannot be equilibrated among all different modes. 
The reason is
that, due to the entropic character of the relaxation, very low energy
modes are rarely involved because the time needed to empty one further
box increases progressively as time goes by,  hence they cannot be thought as
effectively thermalized.  
Note that in the original BG model all boxes
have the same energy, hence there is a unique class of modes. 
For the
general disordered model we expect the existence of a time dependent
energy scale $\eps^*$ separating equilibrated and non-equilibrated
modes.  The mechanism of relaxation is the one we reminded in the
introduction when speaking about the behavior of collective modes.  At
zero temperature there is no thermal activation and the equilibrated
modes are in the sector $\eps\gg \eps^*$ while the non-equilibrated
modes are in the other sector $\eps\ll \eps^*$.  The value of $\eps^*$
can be easily guessed. After quenching to zero temperature the system
starts to relax to its ground state. Because there is no thermal
activation, relaxation is driven by entropic barriers, i.e. flat
directions in configurational space through which the system diffuses.
Entropic relaxation is energy costless so its rate is determined by the
number of available configurations with energy smaller or equal to the
actual energy. A simple microcanonical argument giving the relaxation
rate goes as follows. Let us denote $M$ the number of occupied boxes and
$\Omega(M)$ the number of configurations with $M$ occupied boxes.
The typical time to increase by one unity the number of empty
boxes is given by,

\be
\tau\simeq \frac{\Omega(M)}{\Omega(M-1)}~~~~.
\label{eqBG1}
\ee

For distinguishable particles we have  $\Omega(M)\simeq
M^N$. In the large $N,M$ limit we get
\be \tau\simeq \bigl( \frac{M}{M-1}\bigr )^N\simeq
\exp(\frac{N}{M}) \ .
\label{eqBG2}
\ee

Using the relation $P_0=1-M/N$ and (\ref{eqBG1}) we find
\be
\frac{dP_0}{dt}=-\frac{\Delta M}{N\Delta t}=\frac{1}{\tau}=
\exp\left(-\frac{1}{1-P_0}\right) \ ,
\label{eqBG3}
\ee where we have used the fact that at zero temperature the number of
occupied boxes can only decrease by one unity, thus $\Delta M=-1$.

This yields the result $P_0\simeq 1-\frac{1}{\log(t)+\log(\log(t))}$ in
agreement with the adiabatic approximation.  Actually, from the solution
in (\ref{eqd3}) for $z^*$ and using the adiabatic relation
(\ref{eqd3b}), we obtain $P_0\simeq 1-1/z^*$, which yields the same
result.  The typical relaxation time (\ref{eqBG2}) behaves, then, like
$\tau\simeq \exp(\frac{1}{1-P_0})\simeq \exp(z^*)$.

If the threshold $\eps^*$ plays the role of an energy barrier and
$\beta_{\rm eff}$ accounts for the effective thermal activation due to
entropic effects, we obtain, for the typical relaxation time,
$\tau\simeq \exp(\beta_{\rm eff}\eps^*)$. This expression is only valid
to leading order. As we will see below there are sub-leading corrections
to this expression arising for the fact that the relaxation time is
better described by the expression $\tau\simeq \exp(\beta_{\rm
eff}\eps^*)/(\beta_{\rm eff}\eps^*)$ (see eq. (\ref{eq:taueq})).
Hence, at a given time-scale
$t$ (i.e., the time elapsed since the system was quenched) all modes
where $\tau \ll t $ are equilibrated at zero temperature (which in this
case is the temperature of the thermal bath) and therefore frozen.
Modes with $\tau \gg t$, although dynamically evolving, are also {\em
frozen} because the barriers (in this case entropic barriers) are too
high to allow for relaxation within the time-scale $t$.

Only those modes whose characteristic time is $\tau\sim t$ are relaxing
 at a given time-scale $t$. We get for the time dependent energy scale
 $\eps^*$ and the effective temperature the relation $\eps^*\sim
 \log(t)/\beta_{\rm eff}$ and this yields the leading behavior
 
\be
\eps^*\sim (\log(t))^{-\frac{1}{\alpha+1}}~~~~.
\label{eqe1}
\ee

According to what has been said one can impose the following Ansatz solution
for the  $g_k(\eps)$. If $g_k^{\rm eq}$ stands for the equilibrium
density at $T=0$ (i.e., according to eq. (\ref{eq7g}), 
$g_k^{\rm eq}=g(\eps)\delta_{k,0}$) then we have,

\bea
\Delta g_k(\eps)\equiv g_k(\eps)-g_k^{\rm eq}(\eps)=\frac{\Delta
P_k}{\eps^*}\hat{g}_k\bigl( \frac{\eps}{\eps^*}\bigr)
\label{eqe2}
\eea

\noindent
where $\Delta P_k\equiv P_k-P_k^{\rm eq}=P_k-\delta_{k,0}$ and
$\hat{g}_k(x)$ decays pretty fast to zero for $x > 1$.  This
expression tells us the following. Above $\eps^*$ the $g_k(\eps)$ have
relaxed to their corresponding equilibrium distributions at the
temperature of the bath (in this case the bath is at zero temperature).
On the other hand, in the sector of the energy spectrum where
$\eps<\eps^*$, the densities $g_k$ are still relaxing (specially in the
region $\eps/\eps^*\sim {\cal O}(1)$).  Since the relaxation is driven
by the shift in time of the threshold energy $\eps^*$ the proposed
scaling solution Ansatz seems quite reasonable.  The prefactor $\Delta
P_k/\eps^*$ is introduced to fulfill condition
(\ref{eqPg}). Furthermore the condition $\int_0^\infty dx~\hat
g_k(x)=1$ is imposed on the scaling function $\hat g_k$. 

In the Appendix B we show how this Ansatz  closes the set of equations
(\ref{eqd1}) reproducing  also the leading asymptotic 
behavior for $\eps^*$ and $z^*$ which turns out to be,

\bea
\eps^*\simeq \frac{1}{(\log(t))^{\frac{1}{\alpha+1}}}\label{eqe3}\\
z^*\simeq \frac{1}{(\eps^*)^{1+\alpha}}\simeq \log(t)\label{eqe4}
\eea

For later use we define the following function
\be
G_k(\eps)\equiv\frac{\Delta g_k(\eps)\eps^*}{\Delta P_k}=\hat{g}_k\bigl(
\frac{\eps}{\eps^*}\bigr) \ ,
\label{eqe6}
\ee
\noindent
which scales as function of $\eps/\eps^*$. 
The scaling relation
(\ref{eqe2}) yields the leading asymptotic behavior of all observables
different from the occupation probabilities $P_k$.  
For instance, the
energy is given by $E=-\int_0^{\infty} d\eps~ \eps ~g_0(\eps)$;
using the scaling relation (\ref{eqe2}) and the asymptotic expression
 (\ref{eqe3}) we get for the leading term

\be
E-E_{G S}\sim -\int_0^{\infty} d\eps ~\left[g_0(\eps)-g(\eps)\right]\eps
\sim (\eps^*)^{\alpha+2}\sim 
\frac{1}{\log(t)^{\frac{\alpha+2}{\alpha+1}}}  \ .
\label{eqe5}
\ee

Note that the asymptotic scaling behavior of the energy is the same
as for the effective temperature $T_{\rm eff}=1/\beta_{\rm eff}$ in agreement
with the quasi-equilibrium hypothesis (see (\ref{eqb11})). 
An important result is that the threshold
$\eps^*$ decays slower to zero than the effective temperature. 
A case where this difference can be clearly appreciated corresponds to the case where the
density of states vanishes exponentially fast $g(\eps)\sim
\exp(-A/\eps)$. In this case $\eps^*$ decays slower  than
logarithmically, namely like   $1/\log(\log(t))$ (see appendix B for
details).

\subsection{Relaxational spectrum in equilibrium}

One of the crucial features behind the applicability of the adiabatic
approximation is that the long-time behavior at zero temperature finds
its correspondence with the low-temperature relaxational properties of
the equilibrium state.

To analyze the spectrum of relaxation times $\tau_{\rm eq}(\eps)$ in
equilibrium we expand up to first order in perturbation theory the
dynamical equations for the $g_k(\eps)$ around their equilibrium
solutions $g_k^{\rm eq}(\eps)$. Using the expansion
$g_k(\eps)=g_k^{\rm eq}(\eps)+\delta g_k(\eps)$ we get a set of
equations for the variations $\delta g_k(\eps)$. 
These are shown in the Appendix C.

A complete derivation of the relaxation time $\tau(\eps)$ in
equilibrium is complicated. But it is easy to convince oneself that
the relaxation time is asymptotically (in the limit $T\to 0$) strongly
peaked around the threshold energy $\eps^*$. For $\eps\gg \eps^*$ the
relaxation time is small because the population of high energy boxes
in equilibrium is rather small. On the other hand, for $\eps/\eps^*\ll
1$ the relaxation is estimated to be finite and independent of
$T$. This result is derived in the aforementioned Appendix C where we
show that the maximum relaxation time occurs for $\eps$ around $\eps^*$.
Starting from eqs. (\ref{appC:eqT}) for $\delta g_0(\eps)$ and $\delta
g_1(\eps)$ and making use of the adiabatic Ansatz (\ref{eqe2}), we
find, for $\eps \simeq \eps^*$,
\begin{equation}
\tau(\eps^*)\sim \frac{e^{\beta \eps^*}}{\beta\eps^*}
\label{eq:taueq}
\end{equation}
where $\eps^*(T)\sim T^{1/(2+\alpha)}$, is the asymptotic temperature
dependence of the threshold energy at low temperature. This yields for
the temperature dependence of the relaxation time,

\begin{equation} \tau(T)\sim
\frac{e^{\beta^{\frac{\alpha+1}{\alpha+2}}}}{\beta^{\frac{\alpha+1}{\alpha+2}}}
\label{eq:tauT}
\end{equation}

showing there is activated behavior as a function of the temperature
but with a relaxation time which increases slower than Arrhenius as
$T\to 0$.  Note that for the standard BG model corresponding to
$\alpha\to\infty$ we obtain an Arrhenius behavior and in the opposite
limit $\alpha\to -1$ the relaxation time does not diverge anymore.

\section{Numerical results}

\begin{figure}[tbp]
\begin{center}
{\includegraphics*[width=5.5cm,height=5cm]{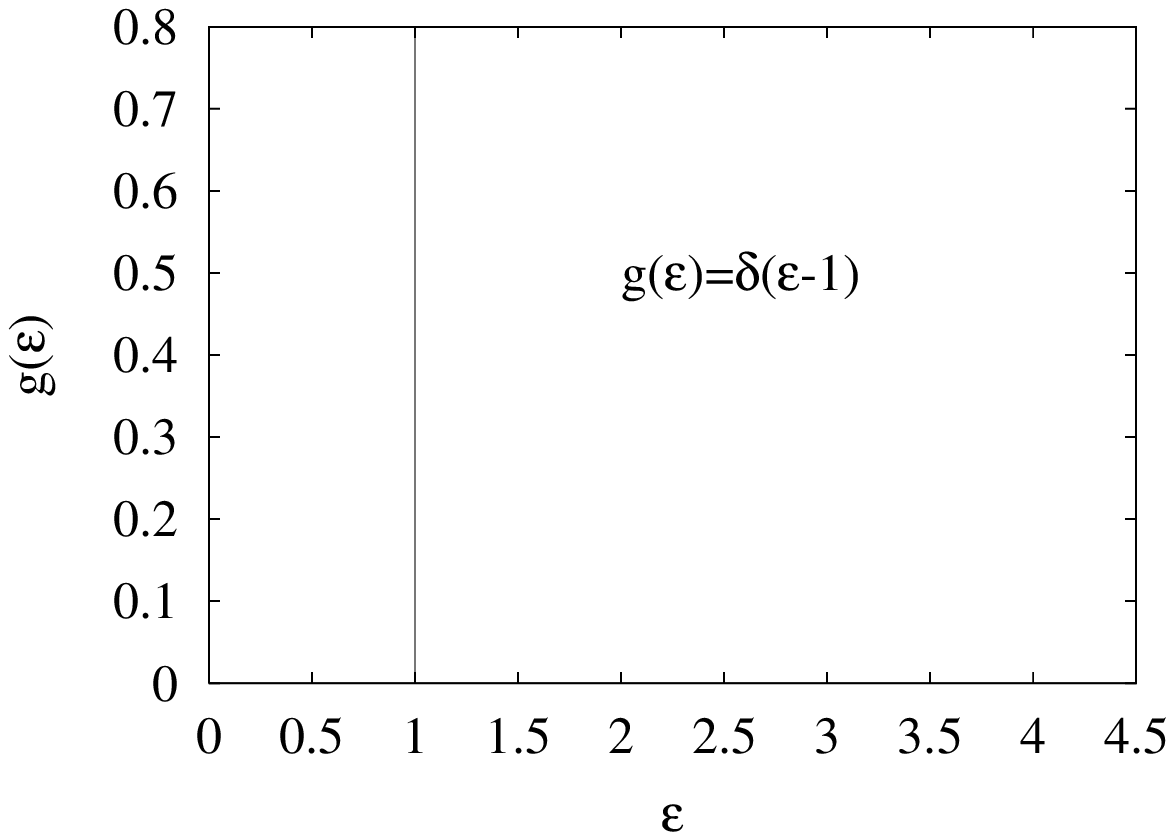}
\includegraphics*[width=5.5cm,height=5cm]{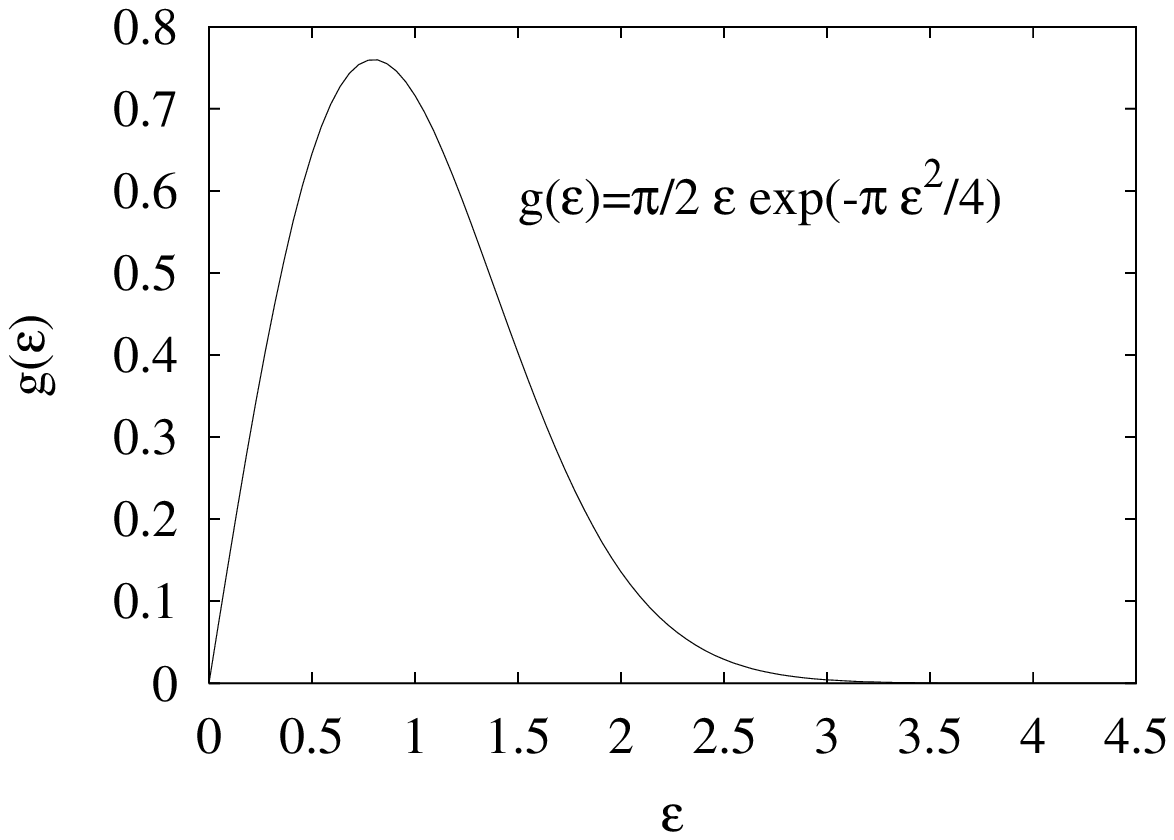}
\includegraphics*[width=5.5cm,height=5cm]{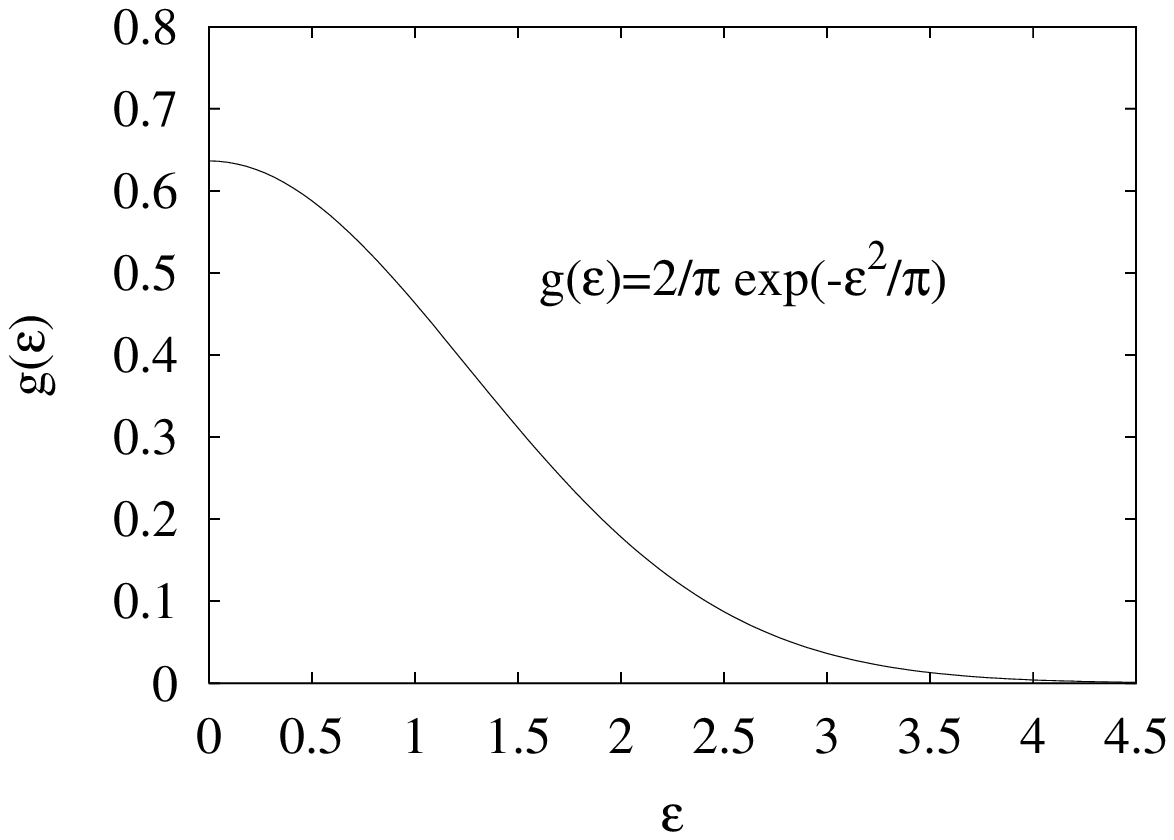}}\vskip 0.05in
\caption{Probability distribution of the energy weights of the boxes of the 
DBG model. (a) The standard Backgammon model has no disordered 
distribution, all boxes have the same weight. (b)
The probability distribution function of a DBG with $\alpha=1$,
at very low energy the density of boxes goes to zero. (c)
A second DBG model with $\alpha=0$. Here the probability of having
boxes with energies arbitrarily close to zero is finite.  \label{fig1}}
\end{center}
\end{figure}

In this section we numerically check the main results obtained in the
previous sections. In particular, we want to show the existence of the
threshold energy $\eps^*$ separating equilibrated from non-equilibrated
energy modes.
We have compared three different models characterized by three different
types of distributions (figure~(\ref{fig1})). All three
distributions were chosen to satisfy the conditions
\be
\int_0^{\infty}d\eps~g(\eps)=\int_0^{\infty}d\eps~\eps~ g(\eps)=1\ , 
\label{eqe7}
\ee
\noindent
in such a way that the ground state has energy $E_{GS}=-1$ in the limit
$N\to\infty$ for all three
cases. The models are the following ones:

\begin{itemize}

\item{{\bf Case A:} {\em Non disordered model with a gap}
(figure~(\ref{fig1}a)).} 
This is the original BG model where
$g(\eps)=\delta(\eps-1)$. 
This case corresponds to $\alpha\to\infty$,
therefore $\eps^*=1$ and the threshold energy is time independent. 
The energy is expected to decay like $E+1\sim T_{\rm eff}\sim 1/\log(t)$.
As previously said in section II, the same behavior is expected for any
disorder distribution $g(\eps)$ with a finite gap.

\item{{\bf Case B:} {\em Disordered Model without gap but $g(0)=0$} 
(figure~(\ref{fig1}b)).}
 We have considered the distribution 
\be
g(\eps)=\frac{\pi}{2}\eps\exp(-\frac{\pi}{4}\eps^2)~~~~.
\label{eqf2}
\ee This case corresponds to $\alpha=1$. The energy threshold $\eps^*$ scales like
$1/\sqrt{\log(t)}$ and the effective temperature and the energy scale
like $E+1\sim T_{\rm eff}\sim 1/(\log(t))^{\frac{3}{2}}$.

\item{{\bf Case C:} {\em Disordered Model without gap and $g(0)$ finite}
(figure~(\ref{fig1}c)). }
We have considered the distribution
\be
g(\eps)=\frac{2}{\pi}\exp(-\frac{\eps^2}{\pi})~~~~.
\label{eqf3}
\ee
This case corresponds to $\alpha=0$. The energy threshold $\eps^*$
scales like $1/\log(t)$ and the effective temperature and the
energy scale like $E+1\sim T_{\rm eff}\sim
1/(\log(t))^{2}$.

\end{itemize}

\begin{figure}[tbp]
\begin{center}
\rotatebox{270}
{\includegraphics*[width=6cm,height=10cm]{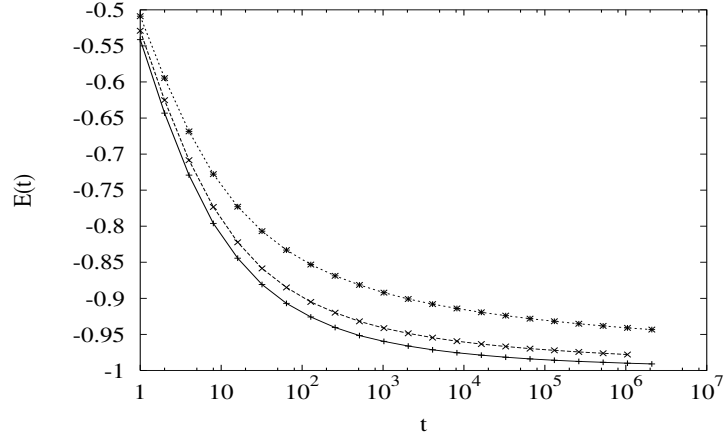}}
\vskip 0.05in
\caption{Energy as a function of time for the three models discussed in
the text. The energy was computed averaging over time intervals
$\Delta t=2^k$, $k$ being an integer number. The lower curve represent
the relaxation in the $\alpha=0$ DBG model (C), the middle curve the
$\alpha=1$ DBG model (B) and th upper curve the standard BG model (A).
\label{fig2}}
\end{center}
\end{figure}

In figure~(\ref{fig2}) we plot the decay of the energy for all three
models. Simulations were done for $N=10^4,10^5,10^6$ boxes (the number
of particles is identical to the number of boxes) showing that
finite-size effects are not big in the asymptotic regime. We show data
for one sample and $N=10^6$. We plot the energy as function of time
starting from a random initial condition (particles randomly
distributed among boxes, $E(t=0)=-1/e$). As clearly seen from the
figure relaxation is faster for the Case C and slower for the standard
BG model (Case A).

\begin{figure}[tbp]
\begin{center}
\rotatebox{270}
{\includegraphics*[width=6cm,height=10cm]{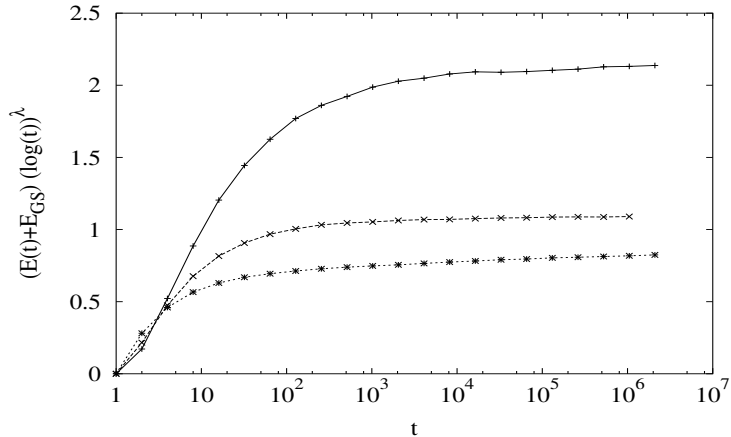}}
\vskip 0.05in
\caption{$(E-E_{GS})(\log(t))^{\lambda}$ plotted as function of time with
$\lambda=\frac{\alpha+2}{\alpha+1}$ for the three different models
discussed in the text. 
The upper curve refers to case C, the middle one to case B and the lower one to standard BG model (case A).
\label{fig3}}
\end{center}
\end{figure}

The different asymptotic behaviors  are shown in
figure~(\ref{fig3}). There we plot
$(E(t)-E_{GS})(\log(t))^{\lambda}$ with
$\lambda\equiv\frac{\alpha+2}{\alpha+1}$. To avoid finite-size corrections
when the energy is close to its ground state we computed exactly
$E_{GS}=\frac{1}{N}(-\sum_{r=1}^N\eps_r+\eps_{\rm min})$ where
$\eps_{\rm min}$ is the minimum value among the $\eps 's$.
The different curves saturate at
a finite quantity corresponding to the asymptotic leading constant. Note
that convergence is slow, showing the presence of subleading
logarithmic corrections to the leading behavior.

\begin{figure}[tbp]
\begin{center}
\rotatebox{0}
{\includegraphics*[width=10cm,height=8cm]{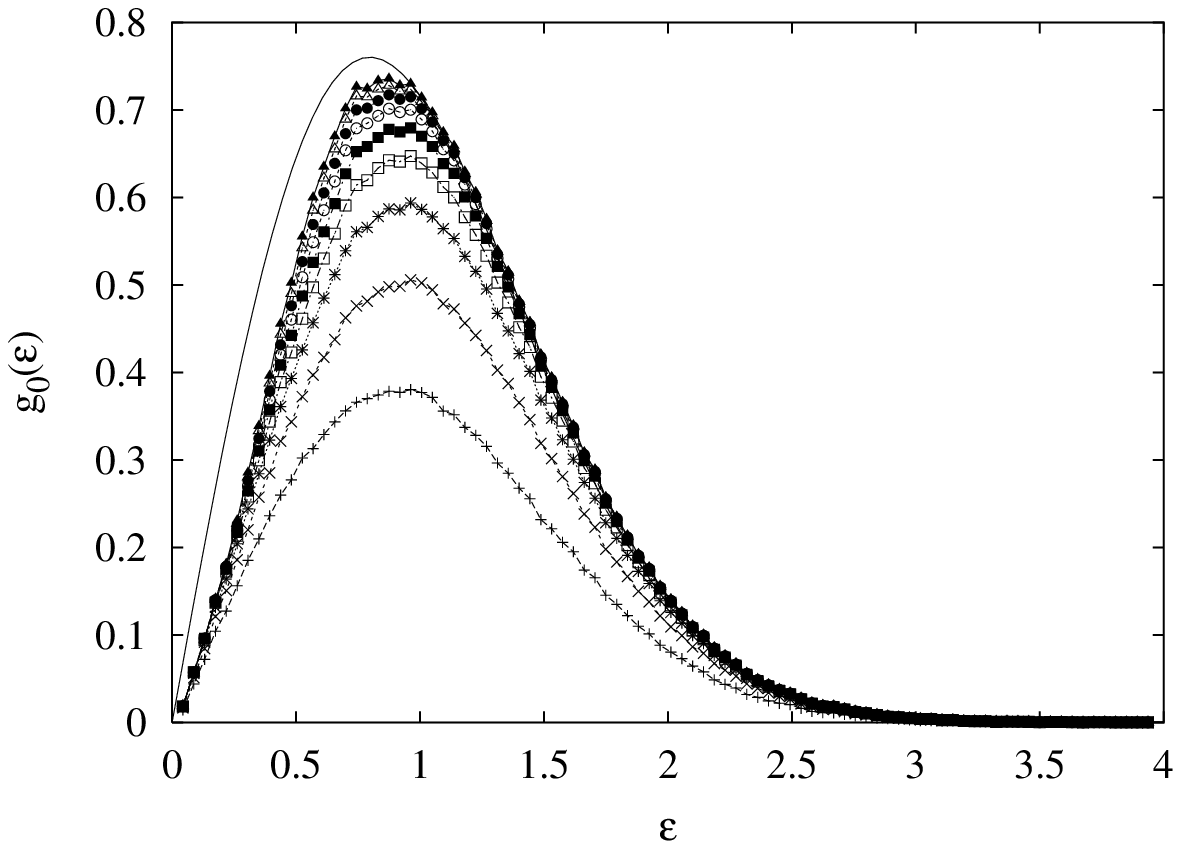}}
\vskip 0.05in
\caption{Distribution $g_0(\eps)$ for case B for different times $2^k$
with $k=4,6,8,10,12,14,16,18,20$ (from bottom to top). The continuous
lines is  $g(\eps)$ given in eq.(\ref{eqf2}).
\label{fig4a}}
\end{center}
\end{figure}

\begin{figure}[tbp]
\begin{center}
\rotatebox{0}
{\includegraphics*[width=10cm,height=8cm]{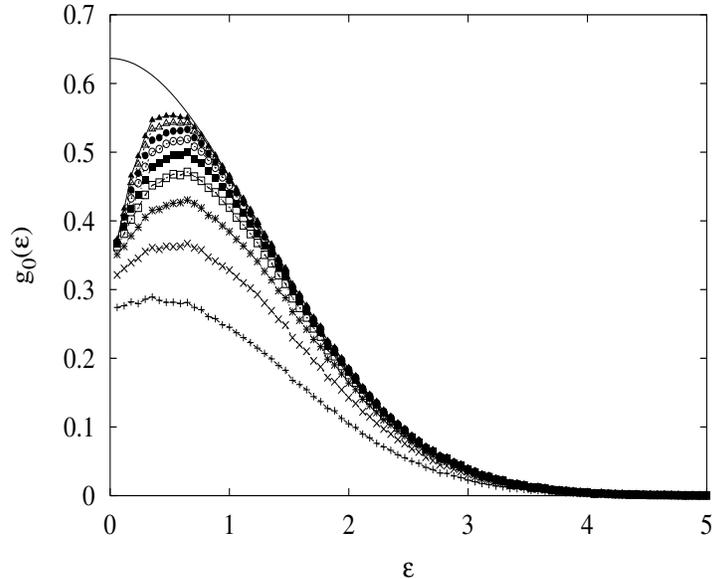}}
\vskip 0.05in
\caption{Distribution $g_0(\eps)$ for case C for different times $2^k$
with $k=4,6,8,10,12,14,16,18,20$ (from bottom to top). The continuous
lines is  $g(\eps)$ given in eq.(\ref{eqf3}).
\label{fig5a}}
\end{center}
\end{figure}

Let us now analyze the shape of the probability densities
$g_k(\eps)$. For these distribution we also only show results for $N=10^6$
because a smaller number of boxes results in higher noise in the
curves. The distribution probabilities were numerically computed by
binning the $\eps$ axis from $\eps=0$ up to $\eps=\eps_{\max}$ where
$\eps_{max}$ is the maximum value of $\eps_r$ among all the $N$
boxes. 
One hundred bins are enough to see the behavior of the time evolution of
the different distributions.  
In figures~(\ref{fig4a}) and (\ref{fig5a})
we show the $g_0(\eps)$ for cases B and C respectively. 
Note that the
$g_0(\eps)$ converge to the asymptotic result $g(\eps)$ for
$\eps>\eps^*$ in agreement with the adiabatic solution (\ref{eqe2})
while they are clearly different for $\eps<\eps^*$. 
The value of
$\eps^*$ where $g_0(\eps)$ deviates from the asymptotic curve $g(\eps)$
shifts slowly to zero (like $1/(\log(t))^{\frac{1}{2}}$ or $1/\log(t)$ for
cases B and C respectively), as can be seen in
figures~(\ref{fig4a},\ref{fig5a})).

\begin{figure}[tbp]
\begin{center}
\rotatebox{0}
{\includegraphics*[width=10cm,height=8cm]{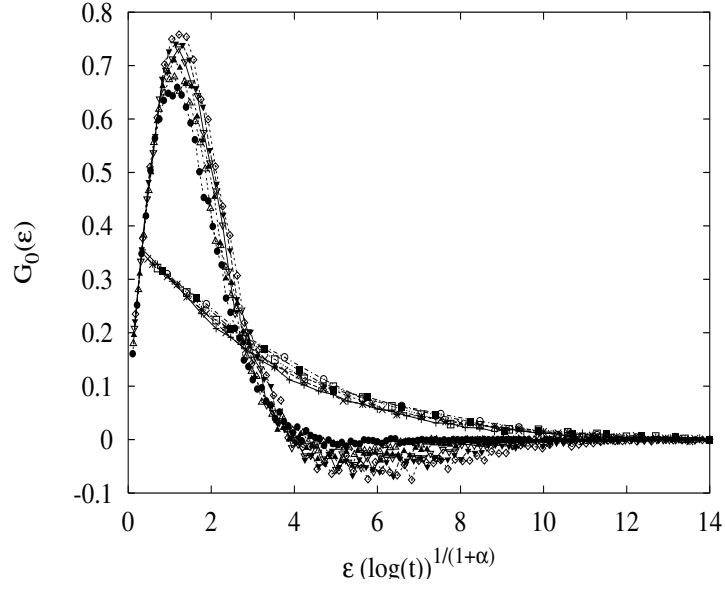}}
\vskip 0.05in
\caption{Distribution $G_0(\eps)$ as function of $\eps\sqrt{\log(t)}$ for
case B (distribution with a maximum) and  as function of $\eps\log(t)$ for
case C (monotonically decreasing distribution). Times are 
$t=2^k$ with $k=6,8,10,12,14,16$. 
\label{fig6a}}
\end{center}
\end{figure}

\begin{figure}[tbp]
\begin{center}
\rotatebox{0}
{\includegraphics*[width=10cm,height=8cm]{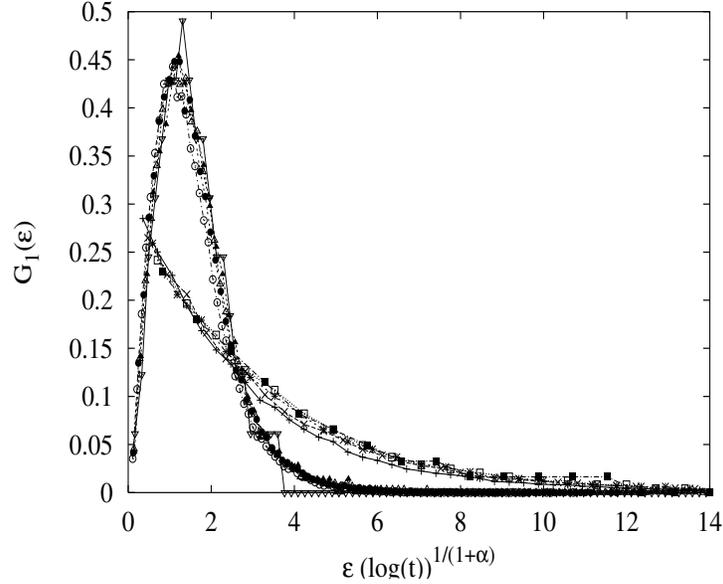}}
\vskip 0.05in
\caption{Distribution $G_1(\eps)$ as function of  $\eps\sqrt{\log(t)}$ for
case B (distribution with a maximum) and  as function of $\eps\log(t)$ for
case C (monotonically decreasing distribution). Times are 
$t=2^k$ with $k=6,8,10,12,14$. 
\label{fig6b}}
\end{center}
\end{figure}

We do not show results for the other $g_k$ (for instance $g_1$) because
they decay very fast to zero (already for $t=2^{17}$ there are no occupied
boxes with more than one particle).  Instead, in figures (\ref{fig6a})
and (\ref{fig6b}) we verify the adiabatic Ansatz,
eqs.(\ref{eqe2},\ref{eqe6}), for the densities $g_0$ and $g_1$ in the two
models B and C. Figure (\ref{fig6a}) plots $G_0(\eps)$ for both
models. Figure (\ref{fig6b}) plots $G_1(\eps)$ for both models.
We have used the relation (\ref{eqe1}) together with
$z^*=\log(t)+\log(\log(t))$ yielding

\bea
&&G_0(\eps)=\frac{\Delta g_0(\eps)\bigl(\log(t)+\log(\log(t))
\bigr)}{\log(t)^
{\frac{1}{\alpha+1}}}=\hat{g}_0\bigl(\frac{\eps}{\eps^*} \bigr)\label{eqf4}\\
&&G_1(\eps)=\frac{\Delta g_1(\eps)t}{\log(t)^
{\frac{\alpha}{\alpha+1}}}=\hat{g}_1\bigl(\frac{\eps}{\eps^*} \bigr)\label{eqf5}
\eea

Note that the scaling is pretty well satisfied and that the $\hat{g}_k(x)$
indeed vanishes for $x\simeq 1$ yielding an estimate for $\eps^*$ is both
cases. We find, $\eps^*\simeq 6/\sqrt{\log(t)}$ for case B and
$\eps^*\simeq 12/\log(t)$ for case C. Note also  that the quality of the
collapse of the $G_0$ is slightly worse for case B than for case C. We
think this is due to the stronger subleading corrections to the shift of
$\eps^*$ which decays slower to zero  for case B. Hence the
asymptotic regime is reached only for later times. Indeed, as figure
(\ref{fig4a}) shows, the value of $\eps^*$ obtained within our time-scales
has  not yet reached the maximum of the distribution $g(\eps)$, so that 
we are
still far from the asymptotic behavior $g(\eps^*)\sim \eps^*$. Yet, it is remarkable how well the scaling Ansatz eqs.(\ref{eqe2},\ref{eqe6}) fits the numerical data.

\section{A method to determine the threshold energy scale $\eps^*$}

In this section we are interested in the following question. Is there
a general method to determine the energy scale $\eps^*$ without having any
precise information about the adiabatic modes present in the system ?
In the previous sections we addressed this question by proposing an
adiabatic scaling Ansatz to the dynamical equations. 
Here we 
propose a general method to determine the energy scale $\eps^*$ from first principles
without the necessity of knowing the nature of the 
slow modes present in the system. Obviously for models such as the standard BG
model this energy scale has no role since we know from the beginning
that relaxation takes place on a single energy scale.

Consider the following quantity $P(\Delta E)$ defined as the
normalized probability density of having a first accepted energy change
$\Delta E$ at time $t$. Let us consider the case of zero temperature
where this probability density is defined only for $\Delta E\le 0$. If
$Q(\Delta E)$ denotes the probability of proposing an energy change at
time $t$ (the move is not necessarily  accepted), it is easy to
show that  $P$ and $Q$ are proportional to each other:

\be
P(\Delta E)=\frac{Q(\Delta E)}{A}\theta(-\Delta E) \ ,
\label{eqPDE}
\ee 
\noindent
where $A=\int_{-\infty}^0Q(\Delta E)d\Delta E$ is the acceptance rate.
The expression for $Q(\Delta E)$ (and therefore $P(\Delta E)$) can be
exactly computed. Note that computing $Q(\Delta E)$ yields all
information about the statistics of energy changes, in particular the
evolution equation for the energy \footnote{Actually, in equilibrium at
finite temperature $Q(\Delta E)$ satisfies detailed balance $Q(\Delta
E)=Q(-\Delta E)\exp(-\beta\Delta E)$.}.  On the contrary, given the time
evolution for the energy this does not necessarily yield the
distribution $Q(\Delta E)$.  For the DBG this function can be exactly
derived (its derivation is shown in the Appendix D). Here we quote the
result,

\be
P(\Delta E)=\frac{\int_0^\infty d\eps~g_0(\eps)~g_1(\eps-\Delta E)
+(1-P_0)g_1(-\Delta E)}{A}\theta(-\Delta E) \ ,
\label{eqPDE0}
\ee
with 
\be
A=\int_0^\infty d\eps'\int_{\eps'}^\infty d\eps~ g_0(\eps')~g_1(\eps)
+(1-P_0)P_1 \ .
\ee

Using the scaling Ansatz eq.(\ref{eqe2}) we obtain the simple scaling 
scaling relation,


\be
P(\Delta E)=\frac{1}{\eps^*}\hat{P}\bigl(\frac{\Delta E}{\eps^*}\bigr )
\label{eqg3}
\ee

\begin{figure}[tbp]
\begin{center}
\rotatebox{270}
{\includegraphics*[width=6cm,height=10cm]{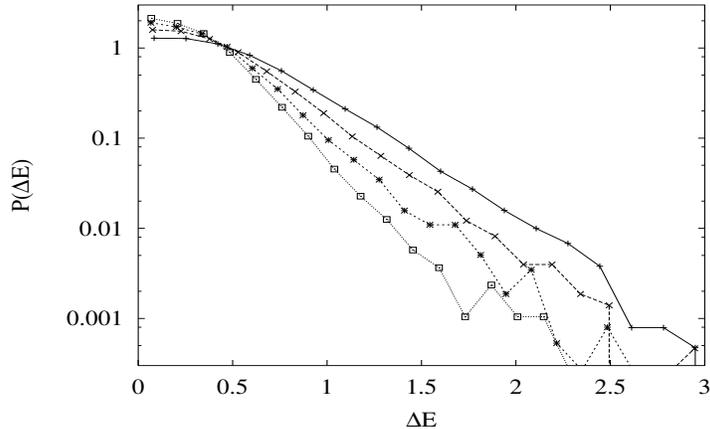}}
\vskip 0.05in
\caption{$P(\Delta E)$ versus $\Delta E$ for different times
$t=10^2,10^3,10^4,10^5$ (from top to bottom) computed as explained in
the text.
\label{fig7}}
\end{center}
\end{figure}

\begin{figure}[tbp]
\begin{center}
\rotatebox{270}
{\includegraphics*[width=6cm,height=10cm]{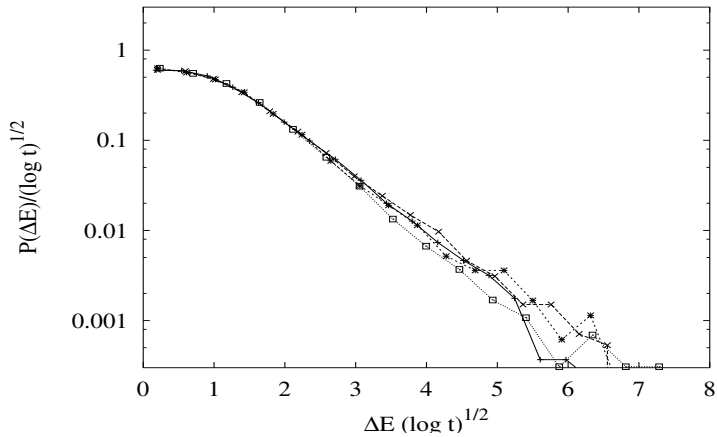}}
\vskip 0.05in
\caption{Scaling plot for $P(\Delta E)/\sqrt{\log(t)}$ versus $\Delta
E\sqrt{\log(t)}$ for different times $t=10^2,10^3,10^4,10^5$.
\label{fig8}}
\end{center}
\end{figure}

A collapse of different $P(\Delta E)$ for different times can be used to
determine the time evolution of $\eps^*$. 
In figure~(\ref{fig7}) we show
the scaling of $P(\Delta E)$ for the model B for $N=10^4$ and
different times $t=10^2,10^3,10^4,10^5$. Starting from a random
initial configuration, statistics has been collected over approximately
$30000$ jumps for every time. In
figure~(\ref{fig8}) we check the scaling relation (\ref{eqg3}) plotting
$P(\Delta E)\eps^*$ as function of $\Delta E/\eps^*$ where we have taken
$\eps^*\sim 1/\sqrt{\log(t)}$. Note also that the range where $P(\Delta
E)$ is finite corresponds to the region where $\eps\sim \eps^*$. In
figure (\ref{fig8}) this corresponds to  $\eps^*\simeq 6/\sqrt{\log(t)}$
in agreement with what was observed in figures (\ref{fig6a},\ref{fig6b}).

The scaling works pretty well showing how this method could be used to
guess the time evolution of the energy threshold $\eps^*$ in general
glassy models in those cases where there exists an energy threshold.

\section{Conclusions}

In this paper we have considered a new solvable glass model, the
disordered Backgammon model (DBG). The new ingredient of this model is
that each box has associated a positive random energy which is obtained
from a distribution $g(\eps)$. Again, similarly to its predecessor (the
Backgammon model (BG)), the model displays slow relaxation due to the
presence of entropic barriers. Actually, it turns out that the
relaxation at $T=0$ of the number of empty boxes and all other
occupation probabilities $P_k$ is exactly the same as the original BG
model, and, in particular,
independent of the disorder distribution $g(\eps)$. In
general, the relaxation of other quantities such as the energy and other
disorder dependent observables, displays an asymptotic relaxation which
depends on the statistical properties of $g(\eps)$ in the limit $\eps\to
0$. In the asymptotic long-time regime, relaxation takes place by
diffusing particles among boxes with the smallest values of
$\eps$. Therefore the asymptotic decay of the energy as well as that of
other observables only depends on the exponent $\alpha$ which
describes the limiting behavior $g(\eps)\to\eps^{\alpha}$. The original BG
model is recovered as a special case in the limit $\alpha\to\infty$.

We have written integral equations for the densities
$g_k(\eps)$. These equations form a hierarchy of dynamical equations
which can be closed by introducing a suitable generating function. We
 focused on the solution of this hierarchy in the particular case
of zero temperature. In this case the analytical solution of these
equations proceeds in two steps. 
Firstly, the equations for the
occupancies $P_k$ are exactly the same as in the original BG model and
they can be solved by using known analytical methods.  Secondly, this
information is used to guess an adiabatic solution for the $g_k(\eps)$
in terms of a time-dependent energy threshold $\eps^*$. 

All densities are out of equilibrium but admit a scaling solution of the type
$\hat{g}_k(\eps/\eps^*)$, with the condition that, for $\eps >\eps^*$,
$\hat{g}_k$ decays to zero fast enough to guarantee that
$\int_0^\infty dx \hat{g}_k(x)=1$. That means that for energies above 
the threshold the modes are almost completely thermalized.

This Ansatz solution yields two types of
leading behaviors: on the one hand, they yield the asymptotic
long-time behavior of $\eps^*,z^*$ at zero temperature; on the other
hand they produce the low-temperature behavior of the values of
$\eps^*,z^*$ in the stationary equilibrium limit $t\to\infty$. 
The
adiabatic approximation is nothing else but stating the validity of
the complementary description of these two very different regimes. 
On
the one hand, the equilibrium regime, where first
the limit $t\to\infty$ is
taken  and later $T\to 0$. 
On the other hand, the far from
equilibrium regime, where the limit $T\to 0$ is taken first and later
$t\to \infty$. The commutation of these two limits allows for the
interchange between different variables such as energy, temperature
and times when expressed in terms of their asymptotic leading
behavior.  Knowing the leading behavior of the quantities $z^*$ and
$\eps^*$,  dimensional reasoning as presented in Appendix B yields the
leading behavior of $\beta_{\rm eff}$ which turns out to be
proportional to $z^*/\eps^*$. The DBG model offers a scenario where
there are two energy sectors separated by the energy scale $\eps^*$
which have very different physical properties. These two
sectors manifest in the behavior of observables like the probability
densities $g_k(\eps)$ where the time dependent threshold $\eps^*$
separates the equilibrated modes ($\eps>\eps^*$) from the modes which
stay off-equilibrium ($\eps<\eps^*$).  In the off-equilibrium regime
entropic barriers are typically higher than the time dependent barrier
at the threshold level $\eps^*$. For $\eps>\eps^*$ barriers are lower
and equilibrium is fastly achieved.

\section{Discussion}

What is the interest of the existence of this energy scale in the
context of real glasses?  One of the most interesting properties of the
present model is that it introduces, in a very simple way, the concept
of a threshold energy scale. To our knowledge such concept has never
been discussed in any one among the plethora of mean-field glassy models
studied during the last years. In those cases one studies relaxation of
global quantities which get contribution of all possible energy scales
involved into the problem. In principle, nothing is wrong with that
since macroscopic observables are those quantities which are always
measured in the laboratory. The problem arises when tackling issues
related to the violation of the fluctuation-dissipation theorem and
concepts such as the effective temperature and partial
equilibration. Usually an effective temperature is defined in terms of
the measured dissipation (response) and fluctuations (correlations) in
the aging regime. This effective temperature is supposed to quantify the
amount of energy transfer when the system is put in contact with a
thermal bath and behaves in several aspects as a real temperature
\cite{TeffCKP,THEO1,GR}. There is a fundamental problem with this definition
which is the following. Suppose one takes a piece of silica well below
$T_g$ (for instance, at room temperature). The piece of silica is not in
equilibrium (actually, it is always relaxing even if the relaxation rate
can be extremely small and unobservable at room temperature) so one
would be tempted to claim that its effective temperature (that
describing equipartition among the set of non-thermalized degrees of
freedom) is around $T_g$ well above the room temperature. Obviously, if
we touch a piece of silica, then the hand plays the role of a thermal
bath at the room temperature. Therefore, why we do not feel the
effective temperature which may be hundreds of degrees above the room
temperature?  Note that the content of energy of the glass is still very
large. Indeed, if the glass suddenly crystallized it would liberate all
its latent heat {\bf{\cite{THEOextra}}}. There are two possible explanations to explain this
discrepancy. The first one was analyzed in the context of the oscillator
model and assumes that the thermal conductivity is so small that heat
transfer is negligible such a short time-scale \cite{GR}. The other
explanation is that, when touching the glass, we are not touching the
slow collective degrees of freedom which still contain a lot of energy
but the thermalized degrees of freedom. These two explanations are not
totally exclusive.  Assuming the existence of an energy threshold
$\eps^*$ such that, above $\eps^*$, all collective modes are thermalized
at the bath temperature and below that threshold they are
off-equilibrium, this offers an explanation about why when touching a
piece of glass we feel it at the room temperature. Our hand only couples
to the higher energy degrees of freedom (phonons) and not to the (much
hotter) collective excitations. Yet, we cannot exclude that even if we
coupled the bath to the hottest collective degrees of freedom then the
conductivity would be extremely small and no heat transfer would be
measured. From a different point of view, the two different explanations
for the small amount of heat transfer established between a bath and the
``hot'' glassy system reduce into a single one: the existence of a
threshold scale $\eps^*$ is consequence of the highly different orders
of magnitude for the conductivities in the two energy sectors. Future
studies of other glassy models will better clarify this issue \cite{LR}.

Finally, we have proposed a method to determine the threshold energy
scale $\eps^*$ by computing the general probability distribution
$Q(\Delta E)$. Preliminary investigations in other glassy models show
that this distribution provides a general way to determine the threshold
scale $\eps^*$. Moreover it
gives interesting information about
fluctuations in the aging state although future work is still needed to
understand better its full implications in our understanding of the
aging regime.

Further investigations in this model will address other issues such as
the measurement of effective temperatures. For instance, it would be
interesting to understand how the effective temperature, defined as the
temperature of the thermal bath which does not produce a net thermal
current when put in contact with the system, depends on the energy
sector $\eps$ probed by the bath. By coupling the bath with a selected 
set of modes of energy $\eps$ we can understand whether there is a single
effective temperature $T_{\rm eff}$ for all modes or rather, there is an $\eps$
dependent temperature. Note that local equilibrium in this model is
only valid in the energy sector $\eps>\eps^*$ and it could well be that there does
not exist a well defined effective temperature in the other sector
$\eps<\eps^*$. Nevertheless, the most natural possibility is that
thermal fluctuations in the off-equilibrium sector $\eps<\eps^*$ are determined 
by the effective temperature (see eq.(\ref{eqd4})) which determines the
relaxation rate of the slow collective mode $P_0$.
Future studies should enlighten this and other related questions.

{\bf Acknowledgments.} We acknowledge A. Garriga and Th. M. Nieuwenhuizen
for a careful reading of the manuscript and suggestions. L.L. is
supported by FOM, The Netherlands.  He acknowledges financial support
from LPTHE where part of this work has been done.  F.R. is supported
PB97-0971 by the Spanish Ministerio de Ciencia y Tecnolog\'{\i}a,
project .  F.R. acknowledge support from the French-Spanish
collaboration (Picasso program and Acciones Integradas HF2000-0097).

\vskip 1cm

\appendix{\bf Appendix A: Occupation probability density equations.}

In this appendix we derive the equations of motion for the occupation 
probability densities for box-energy between $\eps$ and $\eps+d\eps$.
First we start from the densities of having zero particle in a box 
of energy $\eps$.

In the next table we list the processes contributing to the evolution of 
the occupation probability density
of boxes containing 
 zero particle. In the left column we show the processes involved in terms
 of occupation numbers of the departure box and of the arrival box.
In the right column we write the correspondent 
contribution of a given process to 
the variation
of the occupation density, $\Delta g_0(\eps)$.

\begin{center}
\begin{tabular}{|lr|c|}
\hline & {} & {} 
\\
\multicolumn{2}{|c|} {\large{\em occupation}} &
{\large{\em contribution to $\Delta g_0(\eps)$}} 
\\
& {} & {} 
\\
\hline  & {} & {} 
\\
$n_d=1$ & $n_a=0$ &\hspace*{2 mm} $\delta_{n_d,1}\delta_{n_a,0}\left[\delta(\eps-\eps_d)-
\delta(\eps-\eps_a)\right]\left[1+\left(e^{-\beta(\eps_a-\eps_d)}-1\right)
\theta(\eps_a-\eps_d)\right] $\hspace*{2 mm}
\\ 
& {} & {} 
\\
 \cline{2-3}& {} & {} 
\\
& $n_a>0$ &   $\delta_{n_d,1}(1-\delta_{n_a,0})\delta(\eps-\eps_d)$
\\ 
& {} & {} 
\\
 \hline & {} & {} 
\\
$n_d>1$ & $n_a=0$ & 
$-(1-\delta_{n_d,1})\delta_{n_a,0}e^{-\beta\eps_a}\delta(\eps-\eps_a)$
\\  
& {} & {} 
\\
\hline
\end{tabular}
\end{center}

The particle for which a jump is proposed is chosen in box  $d$ 
with probability $n_d/N$. 
The arrival box is chosen with uniform probability $1/N$.

The total  difference per particle in the 
probability density of empty boxes of energy $\eps$ is then

\bea
\Delta g_0(\eps)&=&\frac{1}{N} \sum_{p=0}^N\sum_{a=0}^N\frac{n_p}{N}\frac{1}{N}
\left\{\delta_{n_d,1}\delta_{n_a,0}\left[\delta(\eps-\eps_d)-
\delta(\eps-\eps_a)\right]\left[1+\left(e^{-\beta(\eps_a-\eps_d)}-1\right)
\theta(\eps_a-\eps_d)\right] 
\right.\\
&+&
\left.\delta_{n_d,1}(1-\delta_{n_a,0})\delta(\eps-\eps_d)-
(1-\delta_{n_d,1})\delta_{n_a,0}e^{-\beta\eps_a}\delta(\eps-\eps_a)
\right\} \ .
\eea

Using eqs. (\ref{defP}-\ref{eqPg}) and the following identities,
\bea
&&\frac{1}{N}\sum_{a=0}^N\delta_{n_a,0}\theta(\eps_a-\eps)
\left[e^{-\beta(\eps_a-\eps)}-1\right]=
\int_{\eps}^{\infty}d \eps'~g_0(\eps')~\left[e^{-\beta(\eps'-\eps)}-1
\right] \ ,
\label{appAid1}\\
&&\frac{1}{N}\sum_{d=0}^N n_d \delta_{n_d,1}\theta(\eps-\eps_d)
\left[e^{-\beta(\eps-\eps_d)}-1\right]=
\int_0^{\eps} d \eps'~g_1(\eps')~\left[e^{-\beta(\eps-\eps')}-1
\right] \ ,
\label{appAid2}\\
&&\frac{1}{N}\sum_{a=0}^N\delta_{n_a,0}\delta(\eps-\eps_a)e^{-\beta \eps_a}=
g_0(\eps) e^{-\beta \eps} \ .
\label{appAid3}
\eea
\noindent
we get the equation of motion for $g_0(\eps)$ (namely eq. (\ref{eqg0})):
\bea
\frac{\partial g_0(\eps)}{\partial t}&=&\lim_{N \to \infty} \frac{\Delta g_0(\eps)}{1/N}=
g_1(\epsilon)\left[1+
\int_\epsilon^{\infty}d\epsilon' g_0(\epsilon')\left(
 e^{-\beta(\epsilon'-\epsilon)}-1\right)
\right]
\nonumber
\\
\nonumber
&-&g_{0}(\epsilon) 
\left[e^{-\beta\epsilon}+P_1\left(1-e^{-\beta\epsilon}\right)
+\int_0^\epsilon d\epsilon' g_1(\epsilon')\left(
 e^{-\beta(\epsilon-\epsilon')}-1\right)\right] \ .
\eea

We then consider the evolution of the probability density for 
boxes containing one particle.
In the following table we list the 
processes contributing to the evolution of such occupation probability density.

\begin{center}
\begin{tabular}{|lr|c|}
\hline
\multicolumn{2}{|c|} {\em occupation} &
{\em contribution to $\Delta g_1(\eps)$} 
\\
\hline
$n_d=1$ & $n_a=0$ & $\delta_{n_d,1}\delta_{n_a,0}\left[\delta(\eps-\eps_d)-
\delta(\eps-\eps_a)\right]\left[1+\left(e^{-\beta(\eps_a-\eps_d)}-1\right)
\theta(\eps_a-\eps_d)\right] $
			\\  \cline{2-3}
        & $n_a=1$ &   $-\delta_{n_d,1}\delta_{n_a,1}\left[\delta(\eps-\eps_d)+
\delta(\eps-\eps_a)\right]$
			\\  \cline{2-3}
        & $n_a>1$ &   $-\delta_{n_d,1}(1-\delta_{n_a,1}-\delta_{n_a,0})
\delta(\eps-\eps_d)$
			\\  \hline
$n_d=2$ & $n_a=0$ & $\delta_{n_d,2}\delta_{n_a,0}\left[\delta(\eps-\eps_d)+
\delta(\eps-\eps_a)\right]e^{-\beta\eps_a}$ \\  \cline{2-3}
        & $n_a=1$ &   $\delta_{n_d,2}\delta_{n_a,1}
\left[\delta(\eps-\eps_d)-
\delta(\eps-\eps_a)\right]$ \\  \cline{2-3}
        & $n_a>1$ &   $\delta_{n_d,2}(1-\delta_{n_a,1}-\delta_{n_a,0})
\delta(\eps-\eps_d)$\\  \hline
$n_d>2$ & $n_a=0$ & $(1-\delta_{n_d,2}-\delta_{n_d,1})
\delta_{n_a,0}
\delta(\eps-\eps_a)e^{-\beta\eps_a}$\\  \cline{2-3}
        & $n_a=1$ &   $-(1-\delta_{n_d,2}-\delta_{n_d,1})
\delta_{n_a,1}\delta(\eps-\eps_a)$\\  \hline
\end{tabular}
\end{center}

Departure boxes are chosen
with probability $n_d/N$. 
Arrival box are chosen with uniform probability $1/N$.

Using again eqs. (\ref{defP}-\ref{eqPg}) and eqs. (\ref{appAid1}-\ref{appAid3})
we are able to derive the equation of motion for the probability density 
of boxes with one particles and energy equal to $\eps$:
\bea
\frac{\partial g_1(\eps)}{\partial t}&=&2g_{2}(\epsilon)\left(
1+\int_0^{\infty}d\epsilon g_0(\epsilon) e^{-\beta\epsilon}-P_0\right)
-g_1(\epsilon)\left[2+
\int_\epsilon^{\infty}d\epsilon' g_0(\epsilon')\left(
 e^{-\beta(\epsilon'-\epsilon)}-1\right)
\right]
\nonumber
\\
\nonumber
&+&g_{0}(\epsilon) 
\left[e^{-\beta\epsilon}+P_1\left(1-e^{-\beta\epsilon}\right)
+\int_0^\epsilon d\epsilon' g_1(\epsilon')\left(
 e^{-\beta(\epsilon-\epsilon')}-1\right)\right] \ .
\eea

For densities of boxes with $k>1$ particle the scheme of the contributions
is presented
is the following table:

\begin{center}
\begin{tabular}{|lr|c|}
\hline
\multicolumn{2}{|c|} {\em occupation} &
{\em contribution to $\Delta g_k(\eps)$} 
\\
\hline
$n_d=h<k$ & $n_a=k-1$ & $\delta_{n_d,h}\delta_{n_a,k-1}\delta(\eps-\eps_a)$
\\  
\cline{2-3}
        & $n_a=k$ &   $-\delta_{n_d,h}\delta_{n_a,k}\delta(\eps-\eps_a)
$\\ 
 \hline
$n_d=k$ & $n_a=0$ & $-\delta_{n_d,k}\delta_{n_a,0}\delta(\eps-\eps_a)
e^{-\beta\eps_a}$ \\  \cline{2-3}
        & $0<n_a=h<k-1$ &   $-\delta_{n_d,k}\delta_{n_a,h}\delta(\eps-\eps_a)$ 
\\  \cline{2-3}
        & $n_a=k-1$ &   $-\delta_{n_d,k}\delta_{n_a,k-1}\left[
\delta(\eps-\eps_d)-\delta(\eps-\eps_a)\right]$\\  \cline{2-3}
        & $n_a=k$ &   $-\delta_{n_d,k}\delta_{n_a,k}\left[
\delta(\eps-\eps_d)+\delta(\eps-\eps_a)\right]$\\  \cline{2-3}
        & $n_a>k$ &   $-\delta_{n_d,k}\left(1-\sum_{h=0}^k\delta_{n_a,h}\right)
\delta(\eps-\eps_d)$\\   \hline

$n_d=k+1$ & $n_a=0$ & $-\delta_{n_d,k+1}\delta_{n_a,0}\delta(\eps-\eps_d)e^{-\beta\eps_a}$ \\  \cline{2-3}
        & $0<n_a<k-1$ &   $\delta_{n_d,k+1}\delta_{n_a,h}\delta(\eps-\eps_d)$ \\  \cline{2-3}
        & $n_a=k-1$ &   $\delta_{n_d,k+1}\delta_{n_a,k-1}\left[
\delta(\eps-\eps_d)+\delta(\eps-\eps_a)\right]$\\  \cline{2-3}
        & $n_a=k$ &   $\delta_{n_d,k+1}\delta_{n_a,k}\left[
\delta(\eps-\eps_d)-\delta(\eps-\eps_a)\right]$\\  \cline{2-3}
        & $n_a>k$ &   $\delta_{n_d,k+1}\left(1-\sum_{h=0}^k\delta_{n_a,h}\right)
\delta(\eps-\eps_d)$\\   \hline
$n_d>k$ & $n_a=k-1$ & $\left(1-\sum_{h=1}^{k+1}\delta_{n_d,h}\right)\delta_{n_a,k-1}\delta(\eps-\eps_a)$\\  \cline{2-3}
        & $n_a=k$ &   $-\left(1-\sum_{h=1}^{k+1}\delta_{n_d,h}\right)\delta_{n_a,ky}\delta(\eps-\eps_a)$\\  \hline
\end{tabular}
\end{center}

Combining all the contributions we obtain for $g_k(\eps)$  equation
(\ref{eqgk})
\bea
\frac{\partial g_k(\eps)}{\partial t}&=&(k+1)g_{k+1}(\epsilon)\left(
1+\int_0^{\infty}d\epsilon g_0(\epsilon) e^{-\beta\epsilon}-P_0\right)
\nonumber
\\
\nonumber
&-&g_k(\epsilon)\left[1+k+k\left(
\int_0^{\infty}d\epsilon g_0(\epsilon) e^{-\beta\epsilon}-P_0\right)
\right]+g_{k-1}(\epsilon) \ .
\eea

\vskip 1cm

\appendix{\bf Appendix B: Ansatz for the adiabatic approximation.}


In this appendix we show that the Ansatz solution (\ref{eqe2}) is
asymptotically a solution of the equations (\ref{eqd1}-\ref{eqd1c})
at $T=0$
 yielding the
leading behavior  $\eps^*$ (\ref{eqe3}). We start by rewriting
(\ref{eqe2}) in the following way

\be
\Delta g_k(\eps)=\frac{\Delta
P_k}{\eps} h_k\bigl( \frac{\eps}{\eps^*}\bigr) \ ,
\label{apb1}
\ee
\noindent where 
$\Delta P_k\equiv P_k-\delta_{k,0}$, $
\Delta g_k(\eps)\equiv g_k(\eps)-\delta_{k,0} g(\eps)$,
$h_k(x)=x~\hat{g}_k(x)$ and 
$\int_0^\infty d x ~~ \hat{g}_k(x)=\int_0^\infty d x ~~ h_k(x)/x=1$.
Here we will perform the analysis for the case
$k=0$. The equations for $k>0$ can be done in a similar fashion.
Substituting this expression into eq.(\ref{eqd1}) we get

\bea 
\frac{\partial g_0(\eps)}{\partial t}=
\frac{\partial\Delta P_0
}{\partial t}\frac{1}{\eps}h_0\left(\frac{\eps}{\eps^*}\right)-\frac{\Delta
P_0}{(\eps^*)^2}h_0'(\frac{\eps}{\eps^*})\frac{d\eps^*(t)}{dt}=
-\frac{\Delta P_1}{\eps}h_1(\frac{\eps}{\eps^*})
\Bigl[\int_0^{\eps}d\eps' g(\eps')-\Delta P_0
\int_{\eps}^{\infty}d\eps' \frac{1}{\eps'}
h_0(\frac{\eps}{\eps^*})\Bigr] 
\label{apb2}
\\
\nonumber
+\Delta P_1\left[ g(\eps)+\frac{\Delta
P_0}{\eps}h_0(\frac{\eps}{\eps^*})\int_{\eps}^{\infty}d\eps' \frac{1}{\eps'}
h_1\left(\frac{\eps}{\eps^*}\right)\right]~~~.
\eea
                       \noindent
where $h'_0(x)$ stands for the first derivative of $h_0(x)$.
Note that the scaling function $h_0$ does not depend on time, hence
there is no term $\frac{\partial h_0}{\partial t}$  in that expression. Now
introduce (\ref{eqd2}) in the first term of the l.h.s of (\ref{apb2})
and multiply the whole equation by $\frac{\eps}{\Delta P_0}$ to obtain,

\be 
\Delta P_1~ h_0(x)+x ~h_0'(x)\frac{\partial \log(\eps^*)}{\partial t}
=\Delta P_1~h_1(x)\left[\frac{1}{\Delta P_0}\int_0^{\eps}d\eps'
g(\eps')-\int_{x}^{\infty}dx' \hat{g}_0(x) \right]
- \left[\frac{\eps~
g(\eps)}{\Delta P_0}+h_0(x)\right]\Delta P_1 \int_{x}^{\infty}dx'
\hat{g}_1(x)
\label{apb3}
\ee
\noindent
where $\hat{g}_k(x)=(h_k(x)/x)$. From this equation we can guess the
scaling behavior of all quantities in the asymptotic large-time limit
$\eps^*\to 0$.  In the sector $\eps\le \eps^*$ we use 
$g(\eps)\sim \eps^{\alpha}$
obtaining $\int_0^{\eps}d\eps'g(\eps')\sim \eps^{\alpha+1}$.  Assuming all
terms of the same order, we get for $\eps\sim \eps^*$

\bea
\Delta P_0\sim (\eps^*)^{\alpha+1}\label{apb4}\\
\Delta P_1\sim -\frac{\partial \log(\eps^*)}{\partial t}
\label{apb5}
\eea

Using the standard adiabatic results (\ref{eqd3b}),
$P_0=1-1/z^*,P_1=1/(\exp(z^*))$ we obtain the results
(\ref{eqe3},\ref{eqe4}). Note that the set of equations for $h_k$ are
still impossible to solve. Only in certain regimes such as $\eps\ll \eps^*$ it
may be possible to obtain results. There is a set of equations which
couples the different $h_k$. But this set of equations is time
independent and should yield all the scaling functions $\hat{g}_k(x)$
once appropriate treatment is taking of the amplitude constant which
fixes the leading behavior of $\eps^*$.

We also consider, as an example, the case in which
the probability distribution of
the quenched disorder becomes exponentially high at high values of $\eps$
and zero for low values, namely we choose
\be
g(\eps)=\exp\left(-\frac{A}{\eps}\right) \ .
\ee
For this choice $\int_0^{\eps}d\eps'g(\eps')\sim -\eps 
\exp\left(-\frac{A}{\eps}\right)-A~\Gamma\left(0,\frac{A}{\eps}\right)$, 
where  the generalized Euler
function $\Gamma(0,x)$ goes to zero as $x\to\infty$.
In order to estimate $\eps^*$ from eq. (\ref{apb3}) we notice now that 
for $P_1$ eq. (\ref{apb5}) is still valid, while for 
$\Delta P_0$ we obtain
\be
 \Delta P_0\sim -\eps^*\exp\left(-\frac{A}{\eps^*}\right)\label{apb4b}\ ,
\ee
eventually yielding
\be
\eps^*(t)~\sim~  \frac{A}{\log\left(\log(t)\right)}\ .
\ee

\vskip 1cm

\appendix{\bf Appendix C: Approach to equilibrium of the occupation
densities $g_k(\eps)$.}

We present the equations of motions for the occupation densities
 in the asymptotic 
regime. The values of the densities are expanded to first order 
around their equilibrium
values: $g_k=g_k^{\rm eq} + \delta g_k$.

\bea
\frac{\partial \delta g_0(\eps)}{\partial t}
&=&\delta g_1(\eps)\left\{1+
\int_{\eps}^\infty d\eps' g_0^{\rm eq}(\eps')
\left[e^{-\beta(\eps'-\eps)}-1\right]\right\}
\label{appC:eqT}
\\
\nonumber
&-&\delta g_0(\eps)\left[e^{-\beta\eps}+P_1^{\rm eq}\left(1-e^{-\beta\eps}\right)
+z\int_0^{\eps} d\eps'~g_0^{\rm eq}(\eps') 
\left(e^{-\beta\eps}-e^{-\beta\eps')}\right)\right]
\\
\nonumber&+&g_0^{\rm eq}(\eps)\left\{z\int_\eps^\infty d\eps'~\delta g_0(\eps')
\left(e^{-\beta\eps'}-e^{-\beta\eps}\right)
\right.
\\
\nonumber
&-&\left.\left(1-e^{-\beta\eps}\right)\int_0^\infty d\eps' \delta g_1(\eps') -
\int_0^\eps d\eps'~\delta g_1(\eps')\left[e^{-\beta(\eps-\eps')} -1\right]\right\}
 \hspace*{4 cm}
k=0 \ ,
\\
\frac{\partial \delta g_1(\eps)}{\partial t}&=&
\frac{2}{z}\delta g_2(\eps)-\delta g_1(\eps)\left[2+
\int_{\eps}^\infty d\eps' g_0(\eps') \left[e^{-\beta(\eps'-\eps)}-1\right]
\right]
\\ 
\nonumber&+&\delta g_0(\eps)\left[e^{-\beta\eps} +P_1^{\rm eq}\left(1-e^{-\beta\eps}\right)
+z \int_0^{\eps} d\eps` g_0(\eps') \left(e^{-\beta\eps}-e^{-\beta\eps}\right)\right]
\\
\nonumber&-&g_0^{\rm eq}(\eps)\left\{
z^2 e^{-\beta\eps}\int_0^\infty d\eps'~\delta g_0(\eps')\left(1-e^{-\beta\eps'}\right)
+z\int_\eps^\infty d\eps'~\delta g_0(\eps')
\left(e^{-\beta\eps'}-e^{-\beta\eps}\right)\right.
\\ 
\nonumber&-&\left. \left(1-e^{-\beta\eps}\right)\int_0^\infty d\eps'~\delta g_1(\eps')
-\int_0^\eps d\eps'~\delta g_1(\eps')\left[e^{-\beta(\eps-\eps')}-1\right]
\right\} \hspace*{1 cm}
k=1 \ , 
\\
\frac{\partial \delta g_k(\eps)}{\partial t}&=&
\delta g_{k+1}(\eps)\frac{k+1}{z}-\delta g_{k}(\eps)\left(1+\frac{k}{z}\right) 
+\delta g_{k-1}(\eps)
\\ 
\nonumber&-&
g_{0}^{\rm eq}(\eps)\frac{z^{k+1}}{k!}e^{-\beta\eps}\left(1-\frac{k}{z}\right)
\int_0^\infty
d\eps'~\delta g_0(\eps')~\left(1-e^{-\beta\eps}\right) \ , 
\hspace*{2 cm} k>1 \ .
\eea
In the above equations $\beta$ is the inverse thermal bath temperature and
$z$ is the equilibrium fugacity at that temperature.

As $T$ goes to zero ($\beta \to \infty$, $z(\beta)\to \infty$)
the equations  for the first order perturbation to equilibrium
can be closed:
\bea
\frac{\partial \delta g_0(\eps)}{\partial t}&=&
\delta g_1(\eps)
-\int_{\eps}^\infty d\eps' 
\left[\delta g_1(\eps)~g(\eps') + g(\eps)~\delta g_1(\eps')\right]
\hspace*{2 cm}
k=0 \ ,
\\
\frac{\partial \delta g_1(\eps)}{\partial t}&=&
-2 \delta g_1(\eps)
+\int_{\eps}^\infty d\eps' 
\left[\delta g_1(\eps)~g(\eps') + g(\eps)~\delta g_1(\eps')\right]
\hspace*{2 cm}
k=1 \ ,
\\
\frac{\partial \delta g_k(\eps)}{\partial t}&=&
-\delta g_k(\eps)+\delta g_{k-1}(\eps)  \hspace*{6 cm}  k>1 \ .
\eea

In order to estimate the relaxation characteristic time to equilibrium
at low temperature we can expand eqs. (\ref{appC:eqT}).  First we
introduce the asymptotic threshold energy $\eps^*(T)$ as the energy
discriminating between thermalized and not thermalized collective
modes at temperature $T$. If we define it through the relation
$\eps^*(T)=Tz(T)$ and we use the relation (\ref{eqbz}) obtained by
doing a low T expansion then we get,

\be \eps^*(T)=z_0
T^{\frac{1}{2+\alpha}} \ee where $z_0$ is the coefficient of the
leading term of $z(T)$ at low $T$ (see eq.(\ref{eqbz}): $z(T)=z_0
T^{\frac{1+\alpha}{2+\alpha}}$.

Then we expand eqs. (\ref{appC:eqT}), take $\eps \simeq \eps^*$
and introduce the following adiabatic {\em Ansatz},

 \bea
\delta g_k(\eps)\equiv g_k(\eps)-g_k^{\rm eq}(\eps)=\frac{\Delta
P_k(T,t)}{\eps^*(T)}\hat{g}_k\bigl( \frac{\eps}{\eps^*(T)}\bigr)~~~.
\label{eqe2b}
\eea

Note that this solution is equivalent to the {\em Ansatz}
eq. (\ref{eqe2}) introduced for the asymptotic dynamics at zero
temperature but with a static $\eps^*(T)$ now replacing the dynamical
threshold. Now consider eq.(\ref{appC:eqT}) for $\delta
g_0(\eps)$. Because $\delta P_k=\int d\eps \delta g_k(\eps)$ it can be
shown that the slowest mode corresponds to $k=0$, i.e. $\delta
g_0(\eps)\gg\delta g_k(\eps)$ for $k>0$. Therefore the second term in
the right hand side of eq.(\ref{appC:eqT}) dominates the first and the
second terms. Introducing (\ref{eqe2b}) into eq.(\ref{appC:eqT}) we get
that the relaxation time behaves like,

\be
\tau_{\rm eq}(\eps^*)\propto \frac{e^{\beta\eps^*}}{\beta\eps^*}
\ee

For $\eps >> \eps^*$ the relaxation time is much smaller, since those are the 
modes with lower energy barriers.

\vskip 1cm

\appendix{\bf Appendix D: probability distribution of proposed energy updates.}

\begin{center}
\begin{tabular}{|lr|c|c|}
\hline
\multicolumn{2}{|c|} {\em occupation} &
{\em contribution to $E'-E$} & {\em probability} 
\\
\hline
$n_d=1$ & $n_a=0$ & $-\eps_d+\eps_a$ & $g_1(\eps_d) ~g_0(\eps_a)$
			\\  \cline{2-4}
        & $n_a>0$ &   $-\eps_d$      & $g_1(\eps_d) ~[g(\eps_a)-g_0(\eps_a)]$
			\\  \hline
$n_d>1$ & $n_a=0$ &  $\eps_a$        & $g_0(\eps_a) \frac{1}{N}
\sum_p n_p [g(\eps_d)-g_1(\eps_d)]$
		 \\  \cline{2-4}
	& $n_a>0$ & $0$  	     & $~[g(\eps_a)-g_0(\eps_a)]
 \frac{1}{N}\sum_p n_p [g(\eps_d)-g_1(\eps_d)]~$ \\ \hline
\end{tabular}
\end{center}

The probability distribution $Q(\Delta E)$ of proposed energy updates
is the average of all possible changes, each computed with its probability:
\be
Q(\Delta E)\equiv{\overline{\delta\left(E'-E-\Delta E\right)}}
\ee
where $\Delta E$ is the proposed update, $E$ is the energy of the system before
the updating and $E'$ the energy afterwards.
This means
\bea
Q(\Delta E)&=&\int_0^\infty d \eps\int_0^\infty d\eps' ~g_1(\eps) ~g_0(\eps')~ \delta(\Delta+\eps-\eps')
+\int_0^\infty d \eps\int_0^\infty d\eps' ~g_1(\eps) ~[g(\eps')-g_0(\eps')]
~\delta(\Delta E+\eps)
\nonumber\\
\nonumber
&&\hspace*{5 cm}+\int_0^\infty d \eps\int_0^\infty d\eps'~g_0(\eps') \frac{1}{N}
\sum_p n_p~ [g(\eps)-g_1(\eps)]~\delta(\Delta-\eps')
\\
\nonumber
&&\hspace*{5 cm}+\int_0^\infty d \eps\int_0^\infty d\eps'~[g(\eps')-g_0(\eps')]
 \frac{1}{N}\sum_p ~n_p~ [g(\eps)-g_1(\eps)~]\delta(\Delta E)
\\
&=& \int_{\Delta E}^\infty d \eps~g_0(\eps)~g_1(\eps-\Delta E)
+(1-P_0)~g_1(-\Delta E)~\theta(-\Delta E)-(1-P_1)~g_0(\Delta E)~\theta(\Delta E)
\\
\nonumber
&&\hspace*{9 cm}+(1-P_0)(1-P_1)~\delta(\Delta E)
\eea
 The term with $\delta(\Delta E)$ is the term responsible for diffusive 
motion of particles. Such a contribution does not actually give any 
contribution
to the relaxation of the system and therefore we will not consider it from 
now on.

The probability distribution of accepted energy changes is given by
\be
P(\Delta E)=\frac{Q(\Delta E) W(\beta \Delta E)}{A}
\ee
where $W(\beta \Delta E)$ is the Metropolis function
\be
W(\beta x)=\left\{\begin{array}{ll}
	e^{-\beta x} & \hspace*{0.5 cm} \mbox{if $x>0$} \\
	1	  &    \hspace*{0.5 cm} \mbox{if $x\leq 0$}
	\end{array}
	\right.
\ee

The normalization factor is
\be
A=\int_0^\infty d\eps'\int_{\eps'}^\infty d\eps\left[e^{-\beta(\eps-\eps')}g_0(\eps)~g_1(\eps')+g_0(\eps')~g_1(\eps)\right]+(1-P_0)P_1+(1-P_1)
\int_0^\infty d\eps g_0(\eps)e^{-\beta \eps}
\ee

As $T\to 0$ the distribution $P$ becomes
\be 
P(\Delta E)=\theta(-\Delta E)\frac{\int_0^\infty d\eps~ g_0(\eps)~g_1(\eps-\Delta E)
+(1-P_0)g_1(-\Delta E)}{A}
\ee
with  $A= \int_0^\infty d\eps'~\int_{\eps'}^\infty d\eps~ g_0(\eps')~g_1(\eps)
+(1-P_0)P_1$.

The normalization factor $A$ is actually 
the acceptance rate of the Monte Carlo 
dynamics:
\be
A=\int_{-\infty}^{\infty}dx~W(\beta x)~Q(x)
\ee
as it was defined in \cite{BPR}.

Using the same notation we can write the energy evolution
as 
\bea
\frac{\partial E}{\partial t}&=&
\int_{-\infty}^{\infty}dx~x~W(\beta x)~Q(x)=
-P_1~E
-\int_0^{\infty} d\eps~g_1(\eps)~\eps
\\
\nonumber
&+&\int_0^\infty d\eps'\int_{\eps'}^\infty d\eps~g_1(\eps')~g_0(\eps)
\left[e^{-\beta(\eps-\eps')}-1\right](\eps-\eps')
+(1-P_1)\int_0^\infty d\eps~g_0(\eps)e^{-\beta \eps}\eps\ .
\label{eqE}
\eea
The r.h.s. of this equation can be equivalently  obtained following
the procedure presented in appendix A.
Indeed, by definition of energy density, is
\be
\frac{\partial E}{\partial t}=-\int_0^{\infty} d\eps ~\eps~ 
\frac{\partial g_0(\eps)}{\partial t}
\label{eqE2}
\ee
Inserting eq. (\ref{eqg0}) in eq. (\ref{eqE2}) we get eq. (\ref{eqE}) back.

\end{document}